\documentclass[journal]{IEEEtran}
\ifCLASSINFOpdf
   \usepackage[pdftex]{graphicx}
\else
   \usepackage[dvips]{graphicx}
\fi

\ifCLASSOPTIONcompsoc 
\usepackage[caption=false,font=normalsize,labelfont=sf,textfont=sf]{subfig}
\else
\usepackage[caption=false,font=footnotesize]{subfig}
\fi
\usepackage{stfloats}

\usepackage{mwe,lipsum} 
\usepackage{array}
\usepackage{multirow}  
\usepackage{multicol}
\usepackage{chemarrow}

\usepackage{amsmath}
\usepackage{mdwmath}
\usepackage{amssymb}
\usepackage{cite}

\usepackage[lined,boxed,commentsnumbered, ruled]{algorithm2e}

\allowdisplaybreaks[1]
 \usepackage{changebar}
 \usepackage{color}

\begin{document}
%
\title{Performance Analysis of Low-Interference \emph{N}-Continuous OFDM}
%
%
%


\author{Peng~Wei, Yue~Xiao, Lilin~Dan, Shichao~Lv, and Wei~Xiang
\thanks{
P. Wei is with the Department of Electronic Engineering, Beijing National Research Center for Information Science and Technology, Tsinghua University, Beijing 100084, China (e-mail: wpwwwhttp@163.com).

P. Wei is also with the Tianjin Key Laboratory of Photoelectric Detection Technology and System and the School of Electronics and Information Engineering, Tiangong University, Tianjin 300387, China.

Y. Xiao and L. Dan are with the National Key Laboratory of Science and Technology on Communications, University of Electronic Science and Technology of China, Chengdu, China (e-mail: \{xiaoyue, lilindan\}@uestc.edu.cn).

S. Lv is the School of Cyber Security, University of Chinese Academy of Sciences, Beijing 100049, China and the Beijing Key Laboratory of IoT Information Security Technology, Institute of Information Engineering, Chinese Academy of Sciences, Beijing 100093, China (lvshichao@iie.ac.cn).


W. Xiang is with the College of Science and Engineering, James Cook University, Cairns, Qld. 4878, Australia (e-mail: wei.xiang@jcu.edu.au).
}}%
\maketitle

\begin{abstract}
The low-interference \emph{N}-continuous orthogonal frequency division multiplexing (NC-OFDM) system \cite{Ref24,Ref25} is investigated in terms of power spectrum density (PSD) and bit error rate (BER), to prove and quantify its advantages over traditional NC-OFDM.
The PSD and BER performances of the low-interference scheme are analyzed and compared under the parameters of the highest derivative order (HDO) and the length of the smooth signal.
In the context of PSD, different from one discontinuous point per NC-OFDM symbol in \cite{Ref24}, the sidelobe suppression performance is evaluated upon considering two discontinuous points due to the finite continuity of the smooth signal and its higher-order derivatives.
It was shown that with an increased HDO and an increased length of the smooth signal, a more rapid sidelobe decaying is achieved, for the significant continuity improvement of the OFDM signal and its higher-order derivatives.
However, our PSD analysis also shows that if the length of the smooth signal is set inappropriately, the performance may be degraded, even if the HDO is large.
Furthermore, it was shown in the error performance analysis that under the assumptions of perfect and imperfect synchronization, the low-interference scheme incurs small BER performance degradation for a short length of the smooth signal or a small HDO as opposed to conventional NC-OFDM.
Based on analysis and simulation results, the trade-offs between sidelobe suppression and BER are studied with the above two parameters.



\end{abstract}


\IEEEpeerreviewmaketitle

\section{Introduction}

\IEEEPARstart{O}{rthogonal} frequency division multiplexing (OFDM) has been widely adopted in wireless communications \cite{Ref1, Ref2} since it supports high data transmission rate and is capable of combating the inter-symbol interference (ISI) induced by frequency selective fading channels, such as in enhanced Mobile Broadband (eMBB) scenario  \cite{Ref2,Ref36}.
However, traditional rectangularly pulsed OFDM exhibits severe interference at adjacent channels \cite{Ref3}, since its slow sidelobe decaying in the form of the sinc function leads to high out-of-band (OOB) power emission, which will limit its application in the higher spectral efficiency scenario.

The issue of sidelobe suppression has attracted considerable researches in recent years \cite{Ref4,Ref5,Ref6,Ref10,Ref11,Ref8,Ref7,Ref9}.
Specifically, the classic windowing technique can smooth the edges of OFDM symbols to reduce OOB radiation \cite{Ref4}.
However, its extended guard interval will reduce the spectral efficiency.
Then, aiming for sidelobes close to the band, a class of cancellation carrier techniques were investigated \cite{Ref5,Ref6,Ref7}.
Unfortunately, additional power consumption of cancellation carriers causes a loss in the signal-to-noise ratio (SNR).
Furthermore, the non-orthogonal carriers employed in \cite{Ref6} will result in severe ISI when the range of the targeted band is large, and the reserved carriers may impose an influence on pilot placement \cite{Ref7}.
Recently, precoding methods have been paid more attention based on matrix optimization \cite{Ref10,Ref11,Ref8,Ref9}.
In these methods, the precoders in \cite{Ref10,Ref8,Ref9} in conjunction with frequency diversity may induce high complexity for improving the error performance compared to conventional OFDM, and Ref. \cite{Ref11} provided a low-complexity precoding matrix design, which, however, will induce high interference for a steep OOB spectrum.

In recent years, due to its excellent performance of sidelobe suppression, \emph{N}-continuous OFDM (NC-OFDM) has been widely researched \cite{Ref13,Ref14,Ref15,Ref17,Ref16,Ref19,Ref21,Ref22,Ref20,Ref18,Ref12}.
NC-OFDM was first introduced by Beek \emph{et al.} in \cite{Ref13,Ref12} upon smoothing the OFDM signal and its first \emph{N} derivatives.
Then, it was further refined in \cite{Ref14,Ref15,Ref17,Ref16,Ref19,Ref21,Ref22,Ref20,Ref18} to improve the error performance as opposed to conventional NC-OFDM \cite{Ref13}.
In \cite{Ref14}, via zeroing the beginning and end of each OFDM symbol including its first \emph{N} derivatives, a memoryless precoder was attained at the expense of increased peak-to-average-power ratio (PAPR), while its interference-constrained counterpart is with degraded sidelobe suppression performance.
On the contrary, the optimized percoder in \cite{Ref15} can considerably reduce sidelobes on selected frequencies at the expense of self-interference.
Then, to eliminate the interference, some improved precoders \cite{Ref16,Ref17,Ref18} were designed to facilitate the signal recovery at the receiver at the expense of complexity.
Furthermore, the precoders in \cite{Ref19,Ref20} can effectively improve the error performance by just adding interference to the guard interval.
Nevertheless, the non-cyclic NC-OFDM symbol may be more sensitive to channel-induced ISI and inter-carrier interference (ICI).


Against the aforementioned \emph{N}-continuous precoders designed in the frequency domain, a class of time-domain signal designs with low complexity were proposed for the \emph{N}-continuity in \cite{Ref23,Ref24,Ref25}.
In \cite{Ref23,Ref24}, the concept of a smooth signal linearly combined with basis signals was presented for \emph{N}-continuity.
Furthermore, aiming at reducing the interference, a low-interference scheme was developed with smooth-window truncated basis signals \cite{Ref25}, while maintaining a considerable OOB radiation suppression and low complexity comparable to the counterparts of \cite{Ref19,Ref20}.
However, if the length of the smooth signal is too short, the sidelobe suppression performance of the low-interference scheme will be degraded, even if the highest derivative order (HDO) \emph{N} is big enough.
On the contrary, the excessive long-length smooth signal causes moderate error performance degradation as opposed to conventional OFDM.
Therefore, our objective in this paper is to quantify the performance of the low-interference scheme in terms of the power spectrum density (PSD) and bit error rate (BER), which are regarded as two of the most important performance measures for communication systems.



It is noted that although the average power spectrum expressions were formulated in \cite{Ref14, Ref15,Ref20,Ref21}, they cannot directly show the characteristics of the sidelobe suppression performance with the HDO of \emph{N}.
According to the continuity based principle of sidelobe decaying
presented in \cite{Ref26}, the authors of \cite{Ref24} derived a tight PSD expression, which shows that the sidelobe decays as $f^{-N-2}$.
Note that only one discontinuous point exists by differentiating the OFDM signal during an OFDM symbol period. 
Different from \cite{Ref24}, due to the truncated design of basis signals, two discontinuous points associated with the length and continuity of the smooth signal are active.
Additionally, detailed BER analysis in multipath fading channels for the current \emph{N}-continuous methods is still desired.

Against the above backdrop, the novel contribution of this paper is that, considering the effects of HDO and the length of the smooth signal to the low-interference NC-OFDM scheme, 
we derive an asymptotic PSD expression upon considering the finite continuities at two disjoint points, 
provide a procedure of the BER calculation in the multipath fading channel under the perfect synchronization, 
discuss three exact average signal-to-interference-plus-noise ratio (SINR) expressions of the imperfectly synchronized signal received through the multipath fading channel, and analyze the error performance of time synchronization at the receiver, in the quest for more-compact spectrum and error-resilient transmission in the low-interference scheme.

The remainder of this paper is organized as follows.
In Section II, the low-interference NC-OFDM scheme is introduced.
In Section III, an analysis of the PSD is presented for three cases.
In Section IV, the BER is first evaluated by deriving the closed-form expression of the SINR assuming perfect synchronization, followed by three analyzed SINR expressions associated with symbol time offset (STO) and carrier frequency offset (CFO), and the per-bit SNR at the transmitter and the error variance of time synchronization at the receiver are, respectively, investigated.
Our numerical results are outlined in Section V.
Finally, our conclusions are offered in Section VI.


\begin{figure}
\centering
\parbox{\textwidth}{%
\includegraphics[width=4in]{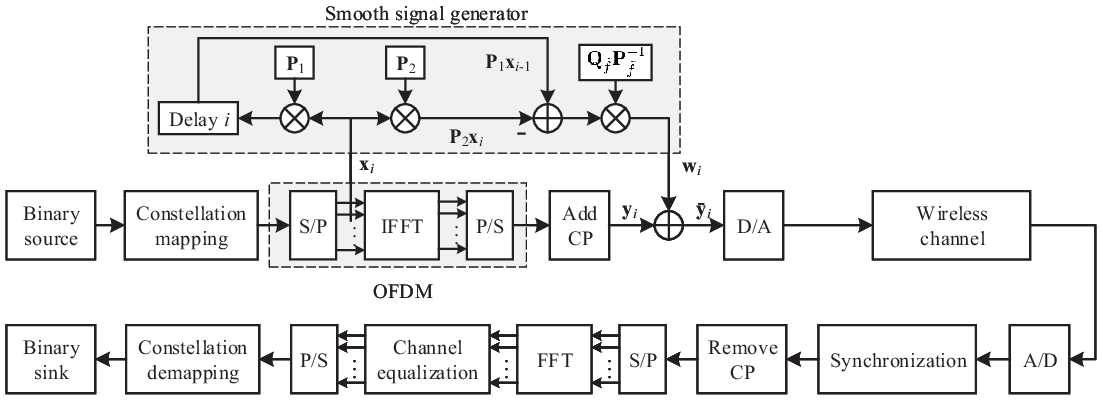}
\DeclareGraphicsExtensions.
\caption{Block diagram of the low-interference {\it N}-continuous OFDM transceiver.}
\label{Fig:Fig1}
}
\end{figure}

\section{Low-Interference NC-OFDM}

The block diagram of the low-interference NC-OFDM transceiver is portrayed in Fig. \ref{Fig:Fig1}.
The binary data stream is first mapped to complex-valued data $x_{i,k_r}$ drawn from phase shift keying (PSK) or quadrature amplitude modulation (QAM).
Then, the $i$th data vector $\mathbf{x}_i=[x_{i,k_0},x_{i,k_1},\ldots,x_{i,k_{K-1}}]^{\rm T}$ is modulated by the OFDM on the subcarrier $\mathcal{K}=\{k_0, k_1, \ldots, k_{K-1}\}$, attached by the cyclic prefix (CP) to generate the $i$th CP-OFDM symbol $y_i(m)$.
Finally, the low-interference scheme employs the time-domain addition of the smooth signal $w_i(m)$ onto the \emph{i}th CP-OFDM symbol $y_i(m)$ as follows \cite{Ref24,Ref25}
\begin{equation}
  \bar{y}_i(m)=y_i(m)+w_i(m),
  \label{Eqn:Eqn1}
\end{equation}
where the sample index $m \in \mathcal{M}=\{-M_{\rm cp}, \ldots, M-1\}$ is with the symbol length $M$ and the CP length $M_{\rm cp}$.
Due to the limited time-domain distribution of $w_i(m)$, whose beginning aligns with the beginning of the $i$th CP-OFDM symbol and tail ends in the front of the $i$th CP-OFDM symbol, to satisfy the \emph{N}-continuity, $w_i(m)$ should follow
\begin{equation}
  \left.{w}^{(n)}_i\!(m)\right|_{m=-M_{\rm cp}}\!\!=\left.y^{(n)}_{i-1}\!(m)\right|_{m=M}
  \!-\left.y^{(n)}_{i}\!(m)\right|_{m=-M_{\rm cp}}\!,
  \label{Eqn:Eqn2}
\end{equation}
where $n \in \mathcal{U}_N  \triangleq \{ 0,1,\ldots,N \}$ and $y^{(n)}_{i}(m)$ is the \emph{n}th derivative of $y_{i}(m)$.


Based on the linear combination of basis signals, the design of $w_i(m)$ is formulated as
\begin{equation}
  {w}_i(m)=\left\{\begin{matrix}
          \sum\limits^{N}_{n=0}{{b}_{i,n}\tilde{f}_n(m)}, & m\in \mathcal{L}, \\
          0, & m\in \mathcal{M}\backslash \mathcal{L},
\end{matrix}\right.
  \label{Eqn:Eqn3}
\end{equation}
where $\mathcal{L}\triangleq\left\{-M_{\rm cp},-M_{\rm cp}\!+\!1,\ldots,-M_{\rm cp}\!+\!L\!-\!1\right\}$ indicates the support length \emph{L} and location of $w_i(m)$.
The basis signals $\tilde{f}_n(m)$ is designed by
\begin{equation}
\tilde{f}_{\tilde{n}}(m)=\left\{\begin{matrix}
          f^{(\tilde{n})}(m)g(m)u(m), & m\in\mathcal{L},  \\
          0, & m\in\mathcal{M}\backslash \mathcal{L},
\end{matrix}\right.
  \label{Eqn:Eqn4}
\end{equation}
which belongs to the basis set $\mathcal{Q}$ defined as
\begin{equation}
  {\mathcal{Q}}\!\triangleq\!\!\left\{\!{\mathbf{q}}_{\tilde{n}}\!\left|{\mathbf{q}}_{\tilde{n}}
  \!=\!\left[\tilde{f}_{\tilde{n}}(-M_{\rm cp}),\ldots,\! \right.   \tilde{f}_{\tilde{n}}(-M_{\rm cp}\!+\!L\!-\!1)\right]^{\rm T}\!\!\!\!, \tilde{n}\!\in \! \mathcal{U}_{2N}\!\right\} \!,
 \label{Eqn:Eqn5}
\end{equation}
where $u(m)$ is the unit-step function, $g(m)$ is a smooth window function with zero edge for the truncation, such as the Hanning, triangular, or Blackman windows, and $f^{(\tilde{n})}(m)$ is expressed by \cite{Ref25}
\begin{equation}
f^{(\tilde{n})}(m)= \left(\frac{j2\pi}{M}\right)^{\tilde{n}}\sum\limits_{k_r\in\mathcal{K}}{k^{\tilde{n}}_r e^{-j\varphi k_r}e^{j2\pi\frac{k_r}{M}m}}.
  \label{Eqn:Eqn6}
\end{equation}

To calculate coefficients $b_{i,n}$, upon solving $N+1$ equations by substituting \eqref{Eqn:Eqn3}-\eqref{Eqn:Eqn6} into \eqref{Eqn:Eqn2}, we have
\begin{equation}
  {\mathbf{b}}_i=\mathbf{P}^{-1}_{\tilde{f}} \left( {\mathbf{P}}_1\mathbf{x}_{i-1}-{\mathbf{P}}_2\mathbf{x}_i  \right),
  \label{Eqn:Eqn7}
\end{equation}
where $\mathbf{b}_i=[b_{i,0},b_{i,1},\ldots,b_{i,N}]^{\rm T}$, $\mathbf{P}_{\tilde{f}}$ is an $(N+1)\times(N+1)$ symmetric matrix, given by
$$\mathbf{P}_{\tilde{f}}=\!\! \begin{bmatrix}
\tilde{f}^{(0)}(-M_{\rm cp}) &\! \tilde{f}^{(1)}(-M_{\rm cp}) &\!  \cdots \!&\!  \tilde{f}^{(N)}(-M_{\rm cp}) \\
\tilde{f}^{(1)}(-M_{\rm cp}) &\! \tilde{f}^{(2)}(-M_{\rm cp}) &\! \cdots \!&\!  \tilde{f}^{(N+1)}(-M_{\rm cp}) \\
\vdots  &\! \vdots  &\! {} \!&\! \vdots\\
\tilde{f}^{(N)}(-M_{\rm cp})  &\! \tilde{f}^{(N+1)}(-M_{\rm cp})  &\! \cdots \!&\!  \tilde{f}^{(2N)}(-M_{\rm cp}) \\
\end{bmatrix}\!,$$
$${\mathbf{P}}_1=
\begin{bmatrix}
1  & 1 & \cdots &  1 \\
\frac{j2\pi k_0}{M}  & \frac{j2\pi k_1}{M} & \cdots & \frac {j2\pi k_{K-1}}{M} \\
\vdots  & \vdots & {} &\vdots\\
\left(\frac{j2\pi k_0}{M}\right)^{N}  &\left(\frac{j2\pi k_1}{M}\right)^{N}  & \cdots &  \left(\frac{j2\pi k_{K-1}}{M}\right)^{N} \\
\end{bmatrix},$$
$\mathbf{P}_{2}={\mathbf{P}}_1\mathbf{\Phi}$, and $\mathbf{\Phi}\triangleq {\rm diag}(e^{j\varphi k_0},e^{j\varphi k_1},\ldots,e^{j\varphi k_{K-1}})$ with $\varphi=-2\pi M_{\rm cp}/M$.


Finally, following the same way of adding $w_i(m)$ in \eqref{Eqn:Eqn1} and the design in \eqref{Eqn:Eqn3} with the preset $\tilde{f}_n(m)$ and the calculated $b_{i,n}$, as shown in Fig. \ref{Fig:Fig1}, the \emph{N}-continuous symbol vector $\bar{\mathbf{y}}_i$ is generated as
\begin{equation}
  \bar{\mathbf{y}}_i=\left\{\begin{matrix}
      \mathbf{y}_i+\left[\!\!\! \begin{array}{c}\mathbf{Q}_{\tilde{f}} \mathbf{P}^{-1}_{\tilde{f}} \left( {\mathbf{P}}_1\mathbf{x}_{i-1}-{\mathbf{P}}_2\mathbf{x}_i  \right) \\ \mathbf{0}_{(M+M_{\rm cp}-L)\times 1}\end{array} \!\!\! \right]\!, & 0\leq i\leq M_{\rm s}-1, \\
      \mathbf{Q}_{\tilde{f}} \mathbf{P}^{-1}_{\tilde{f}} \left( {\mathbf{P}}_1\mathbf{x}_{i-1}-{\mathbf{P}}_2\mathbf{x}_i  \right), &  i=M_{\rm s},
\end{matrix}\right.
  \label{Eqn:Eqn8}
\end{equation}
where $\mathbf{Q}_{\tilde{f}}=[\mathbf{q}_0 \; \mathbf{q}_1 \; \ldots \; \mathbf{q}_N]$ and $M_{\rm s}$ represents the number of successive OFDM symbols. When $i=0$, the initialization of $\mathbf{x}_{i-1}$ is $\mathbf{x}_{-1}=\mathbf{0}_{K\times 1}$.

\section{PSD Analysis}


In continuous time, the transmit \emph{N}-continuous OFDM signal is expressed as
\begin{equation}
  s(t)=
    \sum\limits^{+\infty}_{i=0}\bar{y}_i(t-iT),
  \label{Eqn:Eqn9}
\end{equation}
where $\bar{y}_i(t)$ is the baseband \emph{N}-continuous OFDM symbol, formulated as
\begin{equation}   \label{Eqn:Eqn10}
  \bar{y}_i(t)=\!\begin{cases}
    \sum\limits_{k_r\in \mathcal{K}}\!{\bar{x}_{i,k_r}e^{j2\pi \frac{k_r}{T_{\rm s}}t}},   & -T_{\rm cp}\!\leq\! t \!<\! -T_{\rm cp}\!+\!T_{L}, \\
  \sum\limits_{k_r\in \mathcal{K}}\!{x_{i,k_r}e^{j2\pi \frac{k_r}{T_{\rm s}}t}},    & -T_{\rm cp}\!+\!T_{L}\!\leq\! t \!<\! T_{\rm s},
\end{cases}
\end{equation}%
$T=T_{\rm s}+T_{\rm cp}$ with the symbol period $T_{\rm s}$ and the CP duration $T_{\rm cp}$, $\bar{x}_{i,k_r}$ is the \emph{N}-continuous data on subcarrier $k_r$, and $T_L$ is the duration of the continuous-time smooth signal $w_i(t)$.

Assuming that $s(t)$ and its first \emph{N} derivatives are continuous as well as its (\emph{N}+1)th derivative has finite amplitude discontinuity \cite{Ref28}, the PSD of $s(t)$ can be expressed by \cite{Ref24}
 \begin{align}
 {\Psi}(f) 
    \!=\! \lim\limits_{U\rightarrow \infty}\!{\frac{1}{2UT}E\!\!\left\{\!\left|\sum\limits^{U-1}_{i=-U}{\!\!\frac{\mathcal{F}
   \left\{\! \bar{y}^{(N+1)}_{i}(t) \!\right\}}{\left(j2\pi f\right)^{N+1} e^{j2\pi fiT}} }\right|^2\!\right\}},
   \label{Eqn:Eqn11}
 \end{align}
where $\mathcal{F}\{\cdot\}$ represents the Fourier transform in the interval of $[-T_{\rm cp}, T_{\rm s}]$.

It is noted that in \cite{Ref24}, Eq. \eqref{Eqn:Eqn11} is conditioned that the discontinuity of the (\emph{N}+1)th derivative just appears between $\bar{y}_{i-1}(t)$ and $\bar{y}_i(t)$.
However, by observing the design of basis signals $\tilde{f}_n(m)$ in \eqref{Eqn:Eqn4}, one can obtain that since the zero-edge truncation window $g(t)$ cannot possess infinite continuous derivatives at its end and its derivatives are unrelated to \emph{N}, differentiating $\bar{y}_i(t)$ will yield another discontinuity at the end of the smooth signal $w_i(t)$, where this discontinuity may not have the same derivative order as the one between $\bar{y}_{i-1}(t)$ and $\bar{y}_i(t)$.
In this case, although \eqref{Eqn:Eqn11} is still available for the {\it N}-continuous signal, since it is independent to the length of the smooth signal, it may not completely describe the PSD of the low-interference scheme, depending on the {\it N}-continuity and the length of the smooth signal associated with discontinuities at two different times.
In the following, without loss of generality, the symmetric Blackman window is taken as an example of $g(t)$, where $g(t)=0.42-0.5\cos{(\pi t/T_L)}+0.08\cos{(2\pi t/T_L)}$ for $t\in [0,2T_L]$ with $T_L=(L-1)T_{\rm samp}$ and the sampling interval $T_{\rm samp}=T_{\rm s}/M$.

\begin{figure}[tphb]
\centering
\includegraphics[width=2.5in]{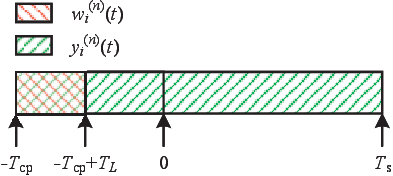}
\DeclareGraphicsExtensions.
\caption{Discontinuous points of higher-order derivatives of the low-interference NC-OFDM signal at times $-T_{\rm cp}$ and $-T_{\rm cp}+T_L$ in a baseband OFDM symbol.}
\label{Fig:Fig2}
\end{figure}

According to the first and second derivatives of the Blackman window function, $0.5(\pi/T_L)\sin(\pi t/T_L)-0.16(\pi/T_L)\sin(2\pi t/T_L)$ and $0.5(\pi/T_L)^2\cos(\pi t/T_L)-0.32(\pi/T_L)^2\cos(2\pi t/T_L)$, we obtain that at the edges of 0 and $2T_L$, the former is continuous with zero while the latter is discontinuous with zero but has finite amplitude.
Furthermore, according to \eqref{Eqn:Eqn4}, the end of $w_i(t)$ corresponds to the edge of $g(t)$ at time $2T_L$.
Thus, as shown in Fig. \ref{Fig:Fig2}, when $N\geq 1$, discontinuities of $\bar{y}^{(N+1)}_i(t)$ may appear at times $-T_{\rm cp}$ or $-T_{\rm cp}+T_L$, which correspond to the beginning and end of $w_i(t)$, respectively.
When $\bar{y}_i(t)$ is differentiated by \emph{N}+1 times, discontinuities at these two disjoint times just simultaneously occur for $N=1$.
Thus, three cases including $N=0$ (e.g. $N<1$), $N=1$, and $N\geq 2$ (e.g. $N>1$)  arise as discussed below.

Additionally, to facilitate the analysis, according to \eqref{Eqn:Eqn3}-\eqref{Eqn:Eqn6} and \eqref{Eqn:Eqn7}, $w_i(t)$ is first written as
\begin{equation}\label{Eqn:Eqn12}
  w_i(t) = \sum\limits^{N}_{n=0}  \sum\limits^{K-1}_{r=0} \left( a_{nr} x_{i-1,k_r} - b_{nr} x_{i,k_r} \right) \tilde{f}_n(t),
\end{equation}
where $a_{nr}$ and $b_{nr}$ are the $(n+1)$th row and $(r+1)$th column elements of $\mathbf{P}^{-1}_{\tilde{f}} {\mathbf{P}}_1$ and $\mathbf{P}^{-1}_{\tilde{f}} {\mathbf{P}}_1\mathbf{\Phi}$, respectively, and $f^{(n)}(t)$ and $g(t)$ consisting of $\tilde{f}_n(t)$ can be expressed as
\begin{equation}
f^{(n)}(t)=
\sum\limits_{k_r\in\mathcal{K}} \left( j2\pi \frac{k_r}{T_{\rm s}} \right)^n e^{j2\pi \frac{k_r}{T_{\rm s}}(t+T_{\rm cp})},
\label{Eqn:Eqn13}
\end{equation}
\begin{align}
g(t) =& 0.42 - 0.5\cos\left( {\pi} (t+T_{\rm cp}+T_L)/{T_L} \right) \nonumber \\
 &+ 0.08\cos\left( {2\pi} (t + T_{\rm cp} + T_L)/{T_L} \right).
\label{Eqn:Eqn14}
\end{align}

\subsection{$N=0$}

It can be shown in Fig. \ref{Fig:Fig2} that by differentiating $s(t)$, the first derivative $\bar{y}^{(1)}_{i}(t)$ at time $-T_{\rm cp}$ and the second derivative $\bar{y}^{(2)}_{i}(t)$ at time $-T_{\rm cp} + T_L$ are both discontinuous.
In this case, the first derivative is considered to evaluate the PSD.
The reasons are explained as follows.
On one hand, since $d_{i,k}$ are independent between two different OFDM symbols, the discontinuity of $\bar{y}^{(1)}_{i}(t)$ at time $-T_{\rm cp}$ is always big in terms of the amplitude and phase, which makes sidelobes decay as $f^{-2}$ based on \eqref{Eqn:Eqn11}.
On the other hand, we have $g(-T_{\rm cp} + T_L)=g^{(1)}(-T_{\rm cp} + T_L)=0$. Since $T_L$ is always much larger than $\pi$ for an effective sidelobe suppression, $g^{(2)}(-T_{\rm cp} + T_L)=0.18(\pi/T_L)^2$ is small.
Hence, with sidelobe decaying of $f^{-3}$ in \eqref{Eqn:Eqn11}, the discontinuity of $\bar{y}^{(2)}_{i}(t)$ is small at time $-T_{\rm cp} + T_L$.
Given the above two reasons, for greater discontinuity and slower sidelobe decaying of $\bar{y}^{(1)}_{i}(t)$ as opposed to $\bar{y}^{(2)}_{i}(t)$, the PSD of $s(t)$ can be expressed as
\begin{equation}
 {\Psi}(f)
    \!=\! \lim\limits_{U\rightarrow \infty}\!{\frac{1}{2UT}E\!\!\left\{\!\left|\sum\limits^{U-1}_{i=-U}{\!\!\frac{\mathcal{F}
   \left\{\bar{y}^{(1)}_{i}(t)\right\}}{j2\pi f}e^{-j2\pi fiT}}\right|^2\!\right\}},
   \label{Eqn:Eqn15}
\end{equation}
where since the support range of $g(t)$ is $[-T_{\rm cp}, -T_{\rm cp}+T_L]$, $\mathcal{F} \left\{\bar{y}^{(1)}_{i}(t)\right\}$ can be expanded as
\begin{align}
\mathcal{F} \! \left\{ \! \bar{y}^{(1)}_{i}(t) \! \right\}
&= \! \frac{1}{T} \!\!\!  \int\limits^{T_{\rm s}}_{-T_{\rm cp}} \!\!\! \!  {y}^{(1)}_{i}(t) e^{-j2\pi ft}dt    \! + \!   \frac{1}{T} \!\!\!\!\!\!\! \int\limits^{-T_{\rm cp}+T_L}_{-T_{\rm cp}} \!\!\!\!\!\!\!\!  {w}^{(1)}_{i}(t)  e^{-j2\pi ft}dt.
\label{Eqn:Eqn16}
 \end{align}

Substituting \eqref{Eqn:Eqn10} and \eqref{Eqn:Eqn12}-\eqref{Eqn:Eqn14} into \eqref{Eqn:Eqn16} yields
\begin{align}
\mathcal{F} \left\{\bar{y}^{(1)}_{i}(t)\right\}
=&\sum\limits^{K-1}_{r=0} \frac{j2\pi k_r}{T_{\rm s}} x_{i,k_r} \frac{{\rm sinc}(f_r (1+\beta_1))}{e^{j\pi f_r (\beta_1 - 1)}}   \nonumber \\
&+ \! \frac{1}{T}\! \sum\limits^{K-1}_{\bar{r}=0} \left( a_{0\bar{r}} x_{i-1,k_{\bar{r}}} \!-\! b_{0\bar{r}} x_{i,k_{\bar{r}}} \right) F_0(f),
   \label{Eqn:Eqn17}
\end{align}
where $\text{sinc}(x)\triangleq \sin(\pi x)/(\pi x)$, $f_r=k_r-T_{\rm s}f$,
\begin{align}
F_0(f)=& 
\sum\limits^{K-1}_{r=0} e^{j\pi(2T_{\rm cp}f + \beta_2 f_r)}  \Bigg( j2\pi k_r \bigg( 0.42 \beta_2 {\rm sinc}(\beta_2 f_r) \nonumber \\
&- \frac{j2\pi f_r \cos(\pi \beta_2 f_r)}{(1-4(\beta_2 f_r)^2) T^{2}_{\rm s}} - \frac{0.08 \beta^{2}_{2} f_r \sin(\pi \beta_2 f_r)}{\pi (1-(\beta_2 f_r)^2)} \bigg) \nonumber \\
&-\frac{\cos(\pi \beta_2 f_r)}{1-4(\beta_2 f_r)^2} + \frac{j 0.16 \sin(\pi \beta_2 f_r)}{1-(\beta_2 f_r)^2} \Bigg),
   \label{Eqn:Eqn18}
\end{align}
$\beta_1=T_{\rm cp}/T_{\rm s}$, and $\beta_2=T_{L}/T_{\rm s}$.

By substituting \eqref{Eqn:Eqn16}-\eqref{Eqn:Eqn18} into \eqref{Eqn:Eqn15}, we have
\begin{align}
 {\Psi}(f)
&= \lim\limits_{U\rightarrow \infty} \frac{1}{2UT} E \Bigg\{ \Bigg|\sum\limits^{U-1}_{i=-U}  e^{-j2\pi fiT}  \nonumber \\
& \cdot \bigg( \sum\limits^{K-1}_{r=0}k_r x_{i,k_r} e^{j\pi f_r (1-\beta_1)} \frac{{\rm sinc}(f_r (1+\beta_1))}{f T_{\rm s}}   \nonumber \\
& + \frac{1}{T} \sum\limits^{K-1}_{\bar{r}=0} \left( a_{0\bar{r}} x_{i-1,k_{\bar{r}}} - b_{0\bar{r}} x_{i,k_{\bar{r}}} \right) \frac{F_0(f)}{j2\pi f} \bigg)
 \Bigg|^2\!\Bigg\},
   \label{Eqn:Eqn19}
\end{align}
which shows that when $s(t)$ is continuous, the PSD of $s(t)$ decays as $f^{-4}$ for a large $f$.

\subsection{$N=1$}

Since the discontinuities of $\bar{y}^{(2)}_{i}(t)$ appear at both $-T_{\rm cp}$ and $-T_{\rm cp}+T_L$, according to \eqref{Eqn:Eqn11}, the PSD of $s(t)$ can be expressed as
\begin{equation}
 {\Psi}(f)
    \!=\! \lim\limits_{U\rightarrow \infty}\!{\frac{1}{2UT}E\!\!\left\{\!\left|\sum\limits^{U-1}_{i=-U}{\!\!\frac{\mathcal{F}
   \left\{\bar{y}^{(2)}_{i}(t)\right\}}{(j2\pi f)^2}e^{-j2\pi fiT}}\right|^2\!\right\}},
   \label{Eqn:Eqn20}
 \end{equation}
where 
\begin{align}
\mathcal{F} \left\{\bar{y}^{(2)}_{i}(t)\right\}
&=\frac{1}{T} \!\!\!  \int\limits^{T_{\rm s}}_{-T_{\rm cp}} \!\!\! {y}^{(2)}_{i}(t) e^{-j2\pi ft}dt   +  \frac{1}{T} \!\!\!\!\!\!\! \int\limits^{-T_{\rm cp}+T_L}_{-T_{\rm cp}} \!\!\!\!\!\!\!\!  {w}^{(2)}_{i}(t)  e^{-j2\pi ft}dt.
 \label{Eqn:Eqn21}
\end{align}

Substituting \eqref{Eqn:Eqn10} and \eqref{Eqn:Eqn12}-\eqref{Eqn:Eqn14} into \eqref{Eqn:Eqn21} yields
\begin{align}
&\mathcal{F} \! \left\{\bar{y}^{(2)}_{i}(t)\right\}
=\sum\limits^{K-1}_{r=0} \!\!  \left( \! \frac{j2\pi k_r}{T_{\rm s}} \! \right)^{\!\! 2} \! x_{i,k_r}  \frac{ {\rm sinc}(f_r (1+\beta_1))}{e^{j\pi f_r (\beta_1-1)}}   \nonumber \\
& \qquad\quad +\! \frac{1}{T} \sum\limits^{1}_{n=0} \! \sum\limits^{K-1}_{\bar{r}=0}\! \left( a_{n\bar{r}} x_{i-1,k_{\bar{r}}} \!-\! b_{n\bar{r}} x_{i,k_{\bar{r}}} \right)\! F_1(n,f),
   \label{Eqn:Eqn22}
\end{align}
where
\begin{align}
F_1(n,&f)
=
\sum\limits^{K-1}_{r=0} \sum\limits^{2}_{\bar{n}=0} \left(\begin{matrix} 2 \\ \bar{n} \end{matrix}\right) \! \left( \! \frac{j2\pi k_r}{T_{\rm s}}\! \right)^{\!\! n+2-\bar{n}}   e^{j\pi (2T_{\rm cp}f+\beta_2 f_r)}   \nonumber \\
&\cdot    \Bigg( 0.42^{(\bar{n})} T_L {\rm sinc}(\beta_2 f_r)   \nonumber \\
& -  \frac{\cos(\pi \beta_2 f_r) \left( \sin(\frac{\pi\bar{n}}{2}) + j2\beta_2 f_r \cos(\frac{\pi\bar{n}}{2}) \right) }{ \left({\pi}/{T_L} \right)^{1- \bar{n}} \left( 1-4(\beta_2 f_r)^2 \right)}    \nonumber \\
&+0.16 \frac{\sin(\pi\beta_2 f_r) \! \left( j\sin(\frac{\pi\bar{n}}{2}) \! - \! \beta_2 f_r \cos(\frac{\pi\bar{n}}{2}) \right)}{ \left({2\pi}/{T_L} \right)^{1- \bar{n}}  \left(1-\left( \beta_2 f_r \right)^2 \right)}\!\! \Bigg) ,
   \label{Eqn:Eqn23}
\end{align}
with the binomial coefficient $\left(\begin{matrix} 2 \\ \bar{n} \end{matrix}\right)$.

It follows from \eqref{Eqn:Eqn21}-\eqref{Eqn:Eqn23} that \eqref{Eqn:Eqn20} can be expressed as
\begin{align}
{\Psi}(f)
&= \lim\limits_{U\rightarrow \infty} \frac{1}{2UT} E \Bigg\{ \Bigg|\sum\limits^{U-1}_{i=-U}  e^{-j2\pi fiT}  \nonumber \\
& \cdot \bigg( \sum\limits^{K-1}_{r=0} k^2_r x_{i,k_r} e^{j\pi f_r (1-\beta_1)} \frac{{\rm sinc}(f_r (1+\beta_1))}{(f T_{\rm s})^2}   \nonumber \\
&-\! \sum\limits^{1}_{n=0}\! \sum\limits^{K-1}_{\bar{r}=0} \!\!\left( a_{n\bar{r}} x_{i-1,k_{\bar{r}}} \!-\! b_{n\bar{r}} x_{i,k_{\bar{r}}} \right) \! \frac{F_1(n,f)}{T (2\pi f)^2} \bigg)\!
 \Bigg|^2\!\Bigg\}.
   \label{Eqn:Eqn24}
\end{align}
It is shown in \eqref{Eqn:Eqn24} that when $s(t)$ and $s^{(1)}(t)$ are both continuous, the PSD of $s(t)$ decays as $f^{-6}$ for large $f$.

\subsection{$N\geq 2$}

Different from the above two cases, upon differentiating $s(t)$ for $N\geq 2$, discontinuity first occurs at time $-T_{\rm cp}+T_L$ and then at time $T_{\rm s}$.
It can be inferred in \eqref{Eqn:Eqn24} that the sidelobe of $s(t)$ decays as at least $f^{-3}$.
Furthermore, since $g(-T_{\rm cp} + T_L)=g^{(1)}(-T_{\rm cp} + T_L)=0$ and when $T_L$ is large $g^{(2)}(-T_{\rm cp} + T_L)=0.18(\pi/T_L)^2$ is small, the discontinuity of $w^{(2)}_i(-T_{\rm cp}+T_L)$ will be small.
Consequently, the sidelobe decaying will approach $f^{-N-2}$, which is mainly determined by the discontinuity at time $-T_{\rm cp}$.

To estimate the effect of the HDO and support length of $w_i(t)$ on the PSD, we introduce another smooth signal $\tilde{w}_i(t)$ to eliminate the discontinuity of $\bar{y}^{(2)}_{i}(t)$ at time $-T_{\rm cp} + T_L$. Thus, according to \eqref{Eqn:Eqn11}, we can derive that
\begin{align}
{\Psi}(f)&=
  \lim\limits_{U\rightarrow \infty} \frac{1}{2UT} E \Bigg\{ \Bigg| \! \sum\limits^{U-1}_{i=-U}\!\! \frac{\mathcal{F} \! \left\{ \! \bar{y}^{(2)}_{i}(t)  \!+\! \tilde{w}_{i}(t) \!-\! \tilde{w}_{i}(t) \! \right\}}{e^{j2\pi fiT} (j2\pi f)^{2}}  \Bigg|^2 \Bigg\} \nonumber \\
   &=  \lim\limits_{U\rightarrow \infty} \frac{1}{2UT} E \Bigg\{ \Bigg|   \sum\limits^{U-1}_{i=-U}  e^{-j2\pi fiT}   \Bigg( - \frac{\mathcal{F} \left\{ \! \tilde{w}_{i}(t) \! \right\}}{(j2\pi f)^2}
   \nonumber \\
   &   + \frac{\mathcal{F} \! \left\{  \left( {y}^{(2)}_{i}(t) + w^{(2)}_{i}(t) + \tilde{w}_{i}(t) \right)^{(N-1)} \right\}}{(j2\pi f)^{N+1}}    \Bigg) \Bigg|^2 \Bigg\}.
   \label{Eqn:Eqn25}
\end{align}

Inspired by \eqref{Eqn:Eqn3}, $\tilde{w}_i(t)$ is also designed by a linear combination of basis signals.
In order to smooth $\bar{y}^{(2)}_{i}(t)=y^{(2)}_{i}(t)+w^{(2)}_{i}(t)$ at time $-T_{\rm cp} + T_L$, the ideal result is that $\bar{y}^{(2)}_{i}(t)+\tilde{w}_i(t)$ possesses infinite continuous derivatives, and the worst result is that $\bar{y}^{(2)}_{i}(t)+\tilde{w}_i(t)$ and its first $N-1$ derivatives are continuous.
According to the design in \eqref{Eqn:Eqn3}, the ideal design requires an infinite number of basis signals.
Therefore, we choose the worse scenario as an example to formulate the following design of $\tilde{w}_i(t)$
\begin{equation}
\tilde{w}_i(t)
=\left\{\begin{matrix}
          \sum\limits^{N-1}_{n=0} \tilde{b}_{i,n} f^{(n)}(t) , & -T_{\rm cp}+T_L  \leq t < T_{\rm s}, \\
          0, & {\rm otherwise}.
          \end{matrix}\right.
  \label{Eqn:Eqn26}
\end{equation}
At the same time, for $\hat{n}=2,3,\ldots,N$, $\tilde{w}_i(t)$ should satisfy
\begin{subequations}  \label{Eqn:Eqn27}
\begin{align}
\left. \tilde{w}^{(\hat{n}-2)}_{i}(t)\right|_{t=T_{\rm s}}&=0,           \label{Eqn:Eqn27a} \\
\left. \tilde{w}^{(\hat{n}-2)}_{i}(t)\right|_{t=-T_{\rm cp}+T_L} &=  \left. w^{(\hat{n})}_{i}(t)\right|_{t=-T_{\rm cp}+T_L}.        \label{Eqn:Eqn27b}
\end{align}
\end{subequations}
By substituting \eqref{Eqn:Eqn26} into \eqref{Eqn:Eqn27}, the coefficients $\tilde{b}_{i,n}$ are calculated by
\begin{equation}
\tilde{\mathbf{b}}_i
=\begin{bmatrix} \mathbf{Q}_1 \\ \mathbf{Q}_2 \end{bmatrix}^{\dag}
\begin{bmatrix} \mathbf{0}_{(N-1)\times 1} \\ \mathbf{w}_{i,L} \end{bmatrix}
=\left( \mathbf{Q}^{\rm H}_1\mathbf{Q}_1 + \mathbf{Q}^{\rm H}_2\mathbf{Q}_2 \right)^{-1}  \mathbf{Q}^{\rm H}_2  \mathbf{w}_{i,L},
  \label{Eqn:Eqn28}
\end{equation}
where $\tilde{\mathbf{b}}_i=[\tilde{b}_{i,0},\ldots,\tilde{b}_{i,N-1}]^{\rm T}$, the Moore-Penrose inverse is
$$\begin{bmatrix} \mathbf{Q}_1 \\ \mathbf{Q}_2 \end{bmatrix}^{\dag}=\left( \mathbf{Q}^{\rm H}_1\mathbf{Q}_1 + \mathbf{Q}^{\rm H}_2\mathbf{Q}_2 \right)^{-1}  \left[\mathbf{Q}^{\rm H}_1 \; \mathbf{Q}^{\rm H}_2\right],$$
$$\mathbf{Q}_1=\begin{bmatrix}
 f^{(0)}(T_{\rm s}) & \cdots & f^{(N-1)}(T_{\rm s}) \\
\vdots & {} & \vdots \\
 f^{(N-2)}(T_{\rm s}) & \cdots & f^{(2N-3)}(T_{\rm s})
\end{bmatrix},$$
$$\mathbf{Q}_2=\begin{bmatrix}
 f^{(0)}(-T_{\rm cp}+T_L) & \cdots & f^{(N-1)}(-T_{\rm cp}+T_L) \\
\vdots & {} & \vdots \\
 f^{(N-2)}(-T_{\rm cp}+T_L) & \cdots & f^{(2N-3)}(-T_{\rm cp}+T_L)
\end{bmatrix},$$
and $\mathbf{w}_{i,L}=[  {w}^{(2)}_i(-T_{\rm cp}+T_L), \ldots, {w}^{(N)}_i(-T_{\rm cp}+T_L) ]^{\rm T}$.

Then, according to \eqref{Eqn:Eqn13}, \eqref{Eqn:Eqn26}, and \eqref{Eqn:Eqn28}, $\tilde{w}_i(t)$ can be expressed by
\begin{align}
\tilde{w}_i(t)=
&
\sum\limits^{N-1}_{n=0} \sum\limits^{N}_{\hat{n}=2} \sum\limits^{K-1}_{r=0} q_{n,\hat{n}} \frac{w^{(\hat{n})}_{i}(T_L\!-\!T_{\rm cp})}{e^{-j2\pi \frac{k_r}{T_{\rm s}} (t+T_{\rm cp})}}  \left( \frac{j2\pi k_r}{T_{\rm s}} \right)^{ n}  ,
  \label{Eqn:Eqn29}
\end{align}
where $q_{n,\hat{n}}$ is the $(n+1)$th row and $(\hat{n}-1)$th column element of matrix $\left( \mathbf{Q}^{\rm H}_1\mathbf{Q}_1 + \mathbf{Q}^{\rm H}_2\mathbf{Q}_2 \right)^{-1}  \mathbf{Q}^{\rm H}_2$ for $\hat{n}=2,3,\ldots,N$ and $n \in \mathcal{U}_{N-1}$. 
Hence, we can obtain that
\begin{align}
 \mathcal{F} & \! \left\{ \tilde{w}^{(N-1)}_{i}(t) \! \right\}= \frac{1}{T}  \!\!\! \int\limits^{T_{\rm s}}_{-T_{\rm cp}+T_L} \!\!\!\!\!\! \tilde{w}_{i}^{(N-1)}(t) e^{-j2\pi ft}dt    \nonumber \\
&\quad =
\frac{T_{\rm s}}{T} \sum\limits^{N-1}_{n=0} \sum\limits^{N}_{\hat{n}=2} \sum\limits^{K-1}_{r=0} q_{n,\hat{n}} w^{(\hat{n})}_{i}(-T_{\rm cp}+T_L)    \nonumber \\
&\quad \quad  \cdot  \left( \frac{j2\pi k_r}{T_{\rm s}} \right)^{N-1+n}   \frac{\sin(\pi(1-\beta_2+\beta_1)f_r)}{\pi f_r  e^{-j\pi ((1+\beta_2)f_r + \beta_1 \bar{f}_r)}},
   \label{Eqn:Eqn30}
\end{align}
where $\bar{f}_r=k_r+T_{\rm s}f$.

Furthermore, we can derive that
\begin{align}
&\mathcal{F} \! \left\{ \! {y}^{(N\!+\!1)}_{i}(t) \! \right\}
\!\!=\! \frac{1}{T} \!\!\!\! \int\limits^{T_{\rm s}}_{-T_{\rm cp}}   {y}^{(N\!+\!1)}_{i}\!(t)   e^{-j2\pi ft}dt   \nonumber \\
&\quad =\! \sum\limits^{K-1}_{r=0} \!\!  \left( \! \frac{j2\pi k_r}{T_{\rm s}}\! \right)^{\!\! N+1} \!\! x_{i,k_r} e^{j\pi f_r (1 - \beta_1)} {\rm sinc}(f_r (1 \!+\! \beta_1)),
   \label{Eqn:Eqn31}
\end{align}
\begin{align}
& \mathcal{F} \! \left\{ {w}^{(N+1)}_{i}(t) \! \right\}= \frac{1}{T}  \!\!\! \int\limits^{-T_{\rm cp}+T_L}_{-T_{\rm cp}} \!\!\!\! {w}_{i}^{(N+1)}(t) e^{-j2\pi ft}dt    \nonumber \\
& \quad =  \frac{1}{T} \! \sum\limits^{N}_{n=0} \sum\limits^{K-1}_{\bar{r}=0} \! \left( a_{n\bar{r}} x_{i-1,k_{\bar{r}}} - b_{n\bar{r}} x_{i,k_{\bar{r}}} \right)  F_2(n,f),
   \label{Eqn:Eqn32}
\end{align}
where $F_2(n,f)$ is given by
\begin{align}
F_2(n,&f)
=
\sum\limits^{K-1}_{r=0} \sum\limits^{N+1}_{\bar{n}=0} \left(\begin{matrix} N\!+\!1 \\ \bar{n} \end{matrix}\right)  \left( \frac{j2\pi k_r}{T_{\rm s}} \right)^{\!\! n-\bar{n}+N+1}     \nonumber \\
& \cdot e^{j\pi (2T_{\rm cp}f+\beta_2 f_r)}  \Bigg(   0.42^{(\bar{n})} T_L   {\rm sinc}\left( \beta_2 f_r \right)    \nonumber \\
&-\cos(\pi \beta_2 f_r)  \frac{\sin(\frac{\pi\bar{n}}{2}) + j2\beta_2 f_r \cos(\frac{\pi\bar{n}}{2})}{(\pi/T_L)^{1-\bar{n}} (1-4(\beta_2 f_r)^2)} \nonumber \\
&+0.16 \sin(\pi \beta_2 f_r)  \frac{ j\sin(\frac{\pi\bar{n}}{2}) - \beta_2 f_r \cos(\frac{\pi\bar{n}}{2})}{(2\pi/T_L)^{1-\bar{n}} (1-(\beta_2 f_r)^2)}   \Bigg).
   \label{Eqn:Eqn33}
\end{align}

Finally, substituting \eqref{Eqn:Eqn29}-\eqref{Eqn:Eqn32} into \eqref{Eqn:Eqn25} yields
\begin{align}
{\Psi}(f)
   &=\lim\limits_{U\rightarrow \infty} \frac{1}{2UT} E \Bigg\{ \Bigg|   \sum\limits^{U-1}_{i=-U}  e^{-j2\pi fiT}   \nonumber \\
   & \cdot \Bigg(  \sum\limits^{K-1}_{r=0}  k^{N+1}_r  x_{i,k_r} e^{j\pi f_r (1 \!-\! \beta_1)} \frac{{\rm sinc}(f_r (1 \!+\! \beta_1))}{(T_{\rm s} f)^{N+1}}  \nonumber \\
   &+\frac{1}{T} \! \sum\limits^{N}_{n=0}\! \sum\limits^{K-1}_{\bar{r}=0} \! \left( a_{n\bar{r}} x_{i-1,k_{\bar{r}}} \!-\! b_{n\bar{r}} x_{i,k_{\bar{r}}} \right) \! \frac{F_2(n,f)}{(j2\pi f)^{N+1}} \nonumber \\
   &- 
   \frac{T_{\rm s}}{T}\! \sum\limits^{N-1}_{n=0} \sum\limits^{N}_{\hat{n}=2} \sum\limits^{K-1}_{r=0} q_{n,\hat{n}} w^{(\hat{n})}_{i}\!(-T_{\rm cp}\!+\!T_L) \! \left( \frac{k_r}{T_{\rm s}} \right)^{\!\! N-1+n}    \nonumber \\
   & \cdot e^{j\pi ((1+\beta_2)f_r + \beta_1 \bar{f}_r)} \frac{\sin(\pi(1-\beta_2+\beta_1)f_r) }{4\pi^3 f_r f^{N+1}   } \nonumber \\
   &+
   \frac{T_{\rm s}}{T} \sum\limits^{N-1}_{n=0} \sum\limits^{N}_{\hat{n}=2} \sum\limits^{K-1}_{r=0} q_{n,\hat{n}} w^{(\hat{n})}_{i}(-T_{\rm cp}\!+\!T_L)  \left(\! \frac{j2\pi k_r}{T_{\rm s}} \! \right)^{\!\!n}  \nonumber \\
   & \cdot e^{j\pi ((1+\beta_2)f_r + \beta_1 \bar{f}_r)} \frac{\sin(\pi(1 \!-\! \beta_2+\beta_1)f_r)}{4\pi^3 f_r f^2}
      \Bigg) \Bigg|^2 \Bigg\}.
   \label{Eqn:Eqn34}
\end{align}

It can be inferred from \eqref{Eqn:Eqn34} that when $f$ is large, the decaying of the PSD is affected by two parts, which have the sidelobe decaying of  $f^{-N-2}$ and $f^{-3}$, respectively.
When $T_L$ increases, since $F_2(n,f)$ in \eqref{Eqn:Eqn33} is in inverse proportion to $T_L$, $F_2(n,f)$ will have a small value.
Furthermore, due to the zero ends of $g(t)$ and $g^{(1)}(t)$ and the small end of $g^{(2)}(t)$ as analyzed above, $w^{(\bar{n})}_i(-T_{\rm cp}+T_L)$ also becomes small.
The above two small values make the last term in \eqref{Eqn:Eqn34} smaller.
Thus, when $f$ is large, as $T_L$ increases, the decaying of the PSD of $s(t)$ approaches $f^{-2N-4}$.
On the contrary, with a reduced $T_L$, the PSD decaying will asymptotically arrive at $f^{-6}$.

\begin{figure}[htp]
\centering
\includegraphics[width=3.5in]{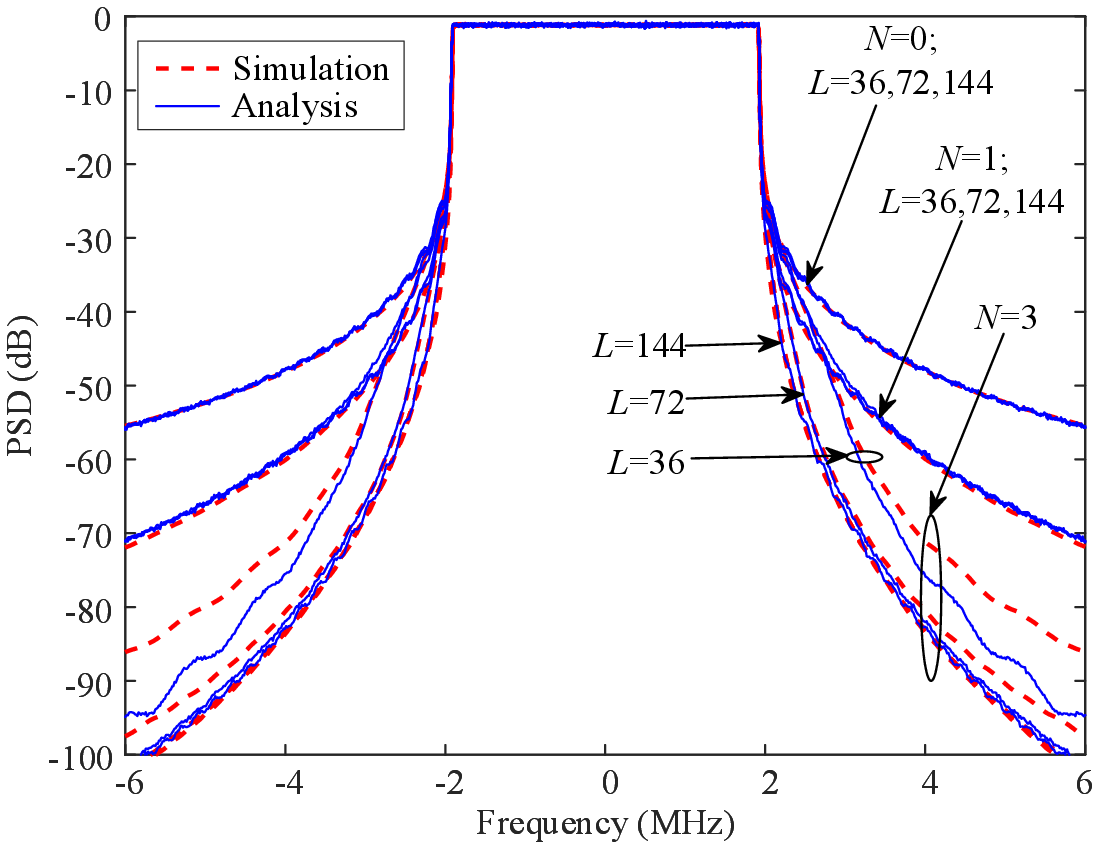}
\DeclareGraphicsExtensions.
\caption{PSD comparison between our analysis and simulation results in low-interference NC-OFDM.}
  \label{Fig:Fig3}
\end{figure}

Fig. \ref{Fig:Fig3} further validates our PSD analysis in the above three cases compared to simulation results with different \emph{L} and $N$, where the parameter configuration will be shown in Section V.
It is shown that the curves of \eqref{Eqn:Eqn19}, \eqref{Eqn:Eqn24}, and \eqref{Eqn:Eqn34} match well with their simulation results.
When $N$ and $L$ increases, the sidelobe decaying speed is significantly enhanced.
On the contrary, decreasing $N$ or $L$ degrades the sidelobe suppression performance.
It is noted that due to the design of \eqref{Eqn:Eqn26} with finite basis signals instead of the ideal design, for $N\geq 2$, there will be a gap between the analytical and simulated results when $f$ is large.

\section{Error Performance Analysis}

In this section, considering perfect synchronization, we first analyze the BER of the low-interference scheme in the multipath fading channel.
Then, with the imperfect frequency and time synchronization at the receiver, the error performance is evaluated in terms of the average SINR.
Thirdly, the power consumption of the smooth signal is investigated in terms of per-bit SNR, e.g., ${\rm E_b/N_0}$.
Finally, based on the correlation function, the error variance of time synchronization at the receiver is analyzed.

\subsection{BER Analysis with Perfect Synchronization}

Through a multipath channel, the \emph{i}th received OFDM symbol $r_i(t)$ is given by
\begin{equation}
  r_i(t)=\sum\limits^{\tilde{L}}_{\tilde{l}=1}{h_{i,\tilde{l}}\bar{y}_i(t-\tau_{\tilde{l}})+n_i(t)}, \; -T_{\rm cp} \leq t < T_{\rm s}
  \label{Eqn:Eqn35}
\end{equation}
where the complex-valued $h_{i,\tilde{l}}$ denotes the time-domain channel coefficient on the $\tilde{l}$th path, following independent and identical distribution with mean zero and variance $\sigma^2_{\tilde{l}}$,
$\tau_{\tilde{l}}$ indicates the time delay along the $\tilde{l}$th path, and the additive white Gaussian noise (AWGN) noise $n_i(t)$ has zero mean and a variance of $\sigma^2_{\rm n}$.

\begin{figure}[htbp]
\centering
\includegraphics[width=3in]{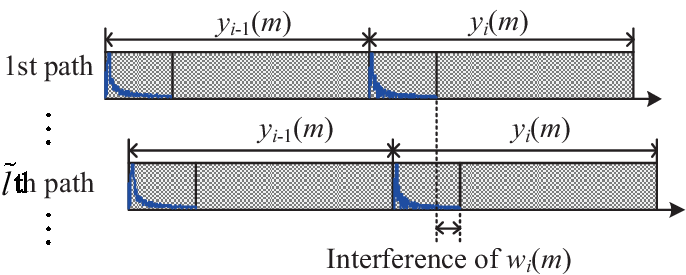}
\DeclareGraphicsExtensions.
\caption{Illustration of the effect of the channel-delayed smooth signal in the multipath fading channel.}
\label{Fig:Fig4}
\end{figure}

Under the assumption of perfect synchronization and a CP capable of combating the adverse channel effect, as shown in Fig. \ref{Fig:Fig1}, after the CP removal followed by an \emph{M}-point discrete Fourier transform (DFT), the received signal on subcarrier $k_r$ becomes
\begin{equation}
  R_{i,k_r}\!=\!H_{i,k_r}\bar{Y}_{i,k_r}\!+\!N_{i,k_r}\!=\!H_{i,k_r}(x_{i,k_r}\!+\!W_{i,k_r})\!+\!N_{i,k_r},
  \label{Eqn:Eqn36}
\end{equation}
where  $R_{i,k_r}$, $\bar{Y}_{i,k_r}$, and $N_{i,k_r}$ are the frequency-domain expressions of $r_i(t)$,  $\bar{y}_i(t)$, and $n_i(t)$, respectively, $H_{i,k_r}$ is the frequency-domain channel response, and $W_{i,k_r}$ is the interference caused by the delayed $w_i(t)$ depicted in Fig. \ref{Fig:Fig4}.

Based on \eqref{Eqn:Eqn36}, the instantaneous SINR on subcarrier $k_r$ can be expressed as
\begin{equation}
  \gamma_{i, k_r}=\dfrac{\left|H_{i,k_r}\right|^2  E\left\{\left|x_{i,k_r} \right|^2\right\}}
  { \left|H_{i,k_r}\right|^2  E\left\{\left|W_{i,k_r} \right|^2\right\}+
  E\left\{ \left|N_{i,k_r}\right|^2\right\}}.
  \label{Eqn:Eqn37}
\end{equation}

It is assumed in \eqref{Eqn:Eqn37} that $x_{i,k_r}$ are independent identically distributed (i.i.d.) random variables with zero mean and unit variance.
Then, we have $E\left\{\mathbf{x}_i\mathbf{x}^{\rm H}_i\right\}=\mathbf{I}_K$ and $E\left\{\mathbf{x}_{i-1}\mathbf{x}^{\rm H}_i\right\}=\mathbf{0}_{K\times K}$.
Moreover, $E\{ |N_{i,k_r}|^2\}$ can be simplified as
\begin{align}
E\!\left\{ \left|N_{i,k_r}\right|^2\right\}\!&=\!E\!\left\{ \mathbf{n}^{\rm H}_{i}\mathbf{F}^{\rm H}_{k_r}\mathbf{F}_{k_r}\mathbf{n}_{i}\right\}\!
={\rm Tr}  \left\{ \mathbf{F}_{k_r}  E\left\{\mathbf{n}_{i}  \mathbf{n}^{\rm H}_{i} \right\}  \mathbf{F}^{\rm H}_{k_r} \right\}   \nonumber \\
&= \frac{\sigma^2_{\rm n}}{M},
  \label{Eqn:Eqn38}
\end{align}
where $\mathbf{F}_{k_r}=\frac{1}{M} [e^{-j2\pi \frac{k_r}{M}0} \; e^{-j2\pi \frac{k_r}{M}} \; \cdots \; e^{-j2\pi \frac{k_r}{M}(M-1)}]$ and $\mathbf{n}_i=[n_i(0),  n_i(1), \ldots, n_i(M-1)]^{\rm T}$.

To calculate $E\{|W_{i,k_r}|^2\}$ in \eqref{Eqn:Eqn37}, the distribution of the smooth signal should be first investigated.
Note that due to the cycle characteristics of basis signals designed in \cite{Ref24}, the interference caused by the smooth signal is independent to the channel delay.
However, based on the design of basis signals in \eqref{Eqn:Eqn4}, the smooth signal $w_i(l)$ is without cycle characteristics.
In this case, as illustrated in Fig. \ref{Fig:Fig4}, varying time delays in different channel paths lead to varying interferences in the delayed tails of ${w}_i(l)$.
With an increased time delay or the length of ${w}_i(l)$, the interference caused by ${w}_i(l)$ increases.
Based on this, we obtain
\begin{align}
& E\left\{\left|W_{i,k_r} \right|^2\right\}  
=E\left\{\left| \mathbf{F}_{k_r}
\sum\limits^{\tilde{L}}_{\tilde{l}=1}  \left[\begin{array}{c} \tilde{\mathbf{w}}_{i,\tilde{l}} \\ \mathbf{0}_{(M-\theta_{\tilde{l}})\times 1} \end{array}\right]        \right|^2\right\} \nonumber \\
&=\! \mathrm{Tr}\!\left\{\!  \mathbf{F}_{k_r}  E\! \left\{\!\!\!  \left(\!\sum\limits^{\tilde{L}}_{\tilde{l}=1} \!\! \left[\!\!\begin{array}{c} \tilde{\mathbf{w}}_{i,\tilde{l}} \\ \mathbf{0}_{(M-\theta_{\tilde{l}})\times 1} \end{array}\!\!\right]\!\!\right)\!\!\!
\left(\! \sum\limits^{\tilde{L}}_{\tilde{l}=1} \!\! \left[\!\!\begin{array}{c} \tilde{\mathbf{w}}_{i,\tilde{l}} \\ \mathbf{0}_{(M-\theta_{\tilde{l}})\times 1} \end{array}\!\!\right]\!\! \right)^{\!\!\!\rm H}\! \right\}\!\! \mathbf{F}^{\rm H}_{k_r}\! \right\}\!,
  \label{Eqn:Eqn39}
\end{align}
where $\tilde{\mathbf{w}}_{i,\tilde{l}}$ denotes the delayed tail of $w_i(l)$ in the $\tilde{l}$th path, expressed by
\begin{equation}
\tilde{\mathbf{w}}_{i,\tilde{l}}=
\begin{bmatrix}
    \mathbf{0}_{\theta_{\tilde{l}}\times(L-\theta_{\tilde{l}})}  &    \mathbf{I}_{\theta_{\tilde{l}}}
\end{bmatrix}
\mathbf{Q}_{\tilde{f}}\mathbf{b}_i,
\label{Eqn:Eqn40}
\end{equation}
and $\theta_{\tilde{l}}={\rm round}\{\tau_{\tilde{l}}/T_{\rm samp}\}+L-M_{\rm cp}$. 
Upon substituting the above equation into \eqref{Eqn:Eqn39}, we have
\begin{align}
E \! \left\{ \! \left|W_{i,k_r} \right|^2 \! \right\}  
& \! =2   \mathbf{F}_{k_r} \mathbf{U} \mathbf{Q}_{\tilde{f}} \mathbf{P}^{-1}_{\tilde{f}} \tilde{\mathbf{P}}_{1} \tilde{\mathbf{P}}^{\rm H}_{1}  \!  \left(\!\mathbf{P}^{-1}_{\tilde{f}}\!\right)^{\rm H}\!  \mathbf{Q}^{\rm H}_{\tilde{f}} \mathbf{U}^{\rm H} \mathbf{F}^{\rm H}_{k_r}    \nonumber \\
& \! =2 \sigma^2_{{\rm w},k_r},
  \label{Eqn:Eqn41}
\end{align}
where the $M\times L$ matrix $\mathbf{U}$ is expressed as
$$\mathbf{U}=\sum\limits^{\tilde{L}}_{\tilde{l}=1}
\left[\begin{array}{c}
\begin{bmatrix}
      \mathbf{0}_{\theta_{\tilde{l}} \times (L-\theta_{\tilde{l}})} & \mathbf{I}_{\theta_{\tilde{l}}}
\end{bmatrix} \\
\mathbf{0}_{(M-\theta_{\tilde{l}})\times L}
 \end{array}\right].$$

Thus, according to \eqref{Eqn:Eqn38} and \eqref{Eqn:Eqn41}, $\gamma_{i, k_r}$ in \eqref{Eqn:Eqn37} reduces to
\begin{equation}
  \gamma_{i, k_r}\!=\!\dfrac{  \left|H_{i,k_r}\right|^2  }
  {2 \sigma^2_{{\rm w},k_r} \left|H_{i,k_r}\right|^2+  \sigma^2_{\rm n}/M }.
  \label{Eqn:Eqn42}
\end{equation}

\begin{figure}[hpt]
\centering
\includegraphics[width=3.5in]{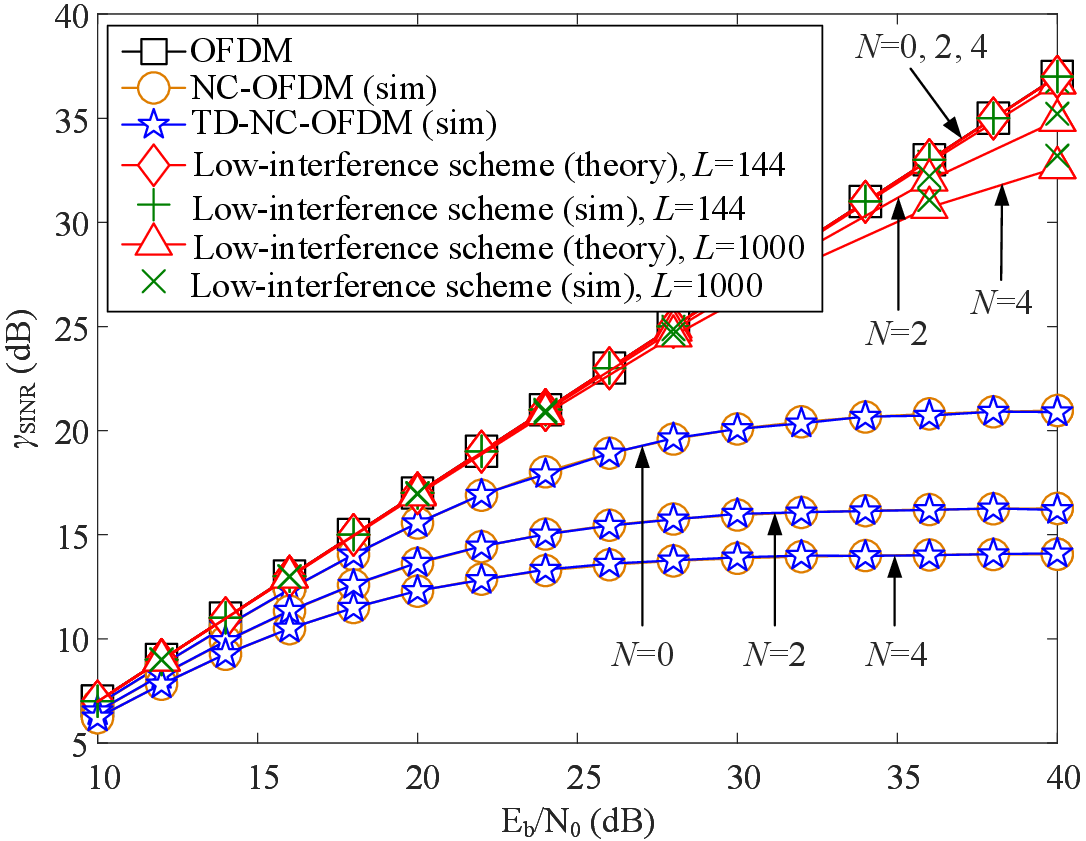}
\DeclareGraphicsExtensions.
\caption{Average SINRs of original OFDM, NC-OFDM, TD-NC-OFDM, and low-interference NC-OFDM with different \emph{L} and \emph{N} in the multipath Rayleigh fading channel.}
\label{Fig:Fig5}
\end{figure}

Fig. \ref{Fig:Fig5} compares the analytical average SINR in \eqref{Eqn:Eqn42} and the simulation results in the 3GPP LTE Extended Vehicular A (EVA) channel model \cite{Ref27}, whose parameters will be given in Section V, where the average SINR is defined as $\gamma_{\rm SINR}=\frac{1}{K}\sum\limits_{k_r\in\mathcal{K}} { E}\{ \gamma_{i, k_r} \}$.
We observe that the analytical $\gamma_{\rm SINR}$ provides a good evaluation to the simulation results.
It also reveals that the SNR loss is negligible in the low-interference NC-OFDM.
Even if the length of $\tilde{w}_i(l)$ becomes longer and the HDO increases, such as $N=4$ and $L=1000$, the low-interference scheme still performs closely to conventional OFDM in the sense of SINR, and is much better than conventional NC-OFDM \cite{Ref13} and time-domain NC-OFDM (TD-NC-OFDM) \cite{Ref24}.

In the following BER analysis, the subscripts denoting the subcarrier and time index are removed for brevity. Hence, by defining $|H_{i,k_r}|^2=\alpha$, $\gamma_{i,k_r}=\gamma$, and $\sigma^2_{{\rm w},k_r}=\sigma^2_{{\rm w}}$, Eq. \eqref{Eqn:Eqn42} can be simplified as
\begin{equation}
  \gamma=\dfrac{ \alpha  } {2 \sigma^2_{\rm w} \alpha   +
  \sigma^2_{\rm n}/M }.
  \label{Eqn:Eqn43}
\end{equation}
Eq. \eqref{Eqn:Eqn43} shows that $\gamma$ is a monotonically increasing continuous derivable function having an inverse $\alpha=f(\gamma)={ (\sigma^2_{\rm n}/M) \gamma}/{(1-2\sigma^2_{\rm w}\gamma)}$. Then, based on the property of the composite function, the probability density function (PDF) $p_{\gamma}(\gamma)$ of $\gamma$ can be formulated as
\begin{equation}
  p_{\gamma}(\gamma)=p_{\alpha}(f(\gamma)) f^{(1)}(\gamma),
  \label{Eqn:Eqn44}
\end{equation}
where the PDF of $\alpha$ is an exponential function $p_{\alpha}(\alpha)=\frac{1}{ E\{\alpha\} } e^{-\frac{\alpha}{{ E}\{\alpha\}}}$ with $E \{ \alpha \}=\sum\limits^{\tilde{L}}_{\tilde{l}=1} \sigma^{2}_{\tilde{l}}$. Hence, Eq. \eqref{Eqn:Eqn44} becomes
\begin{equation}
  p_{\gamma}(\gamma)\!=\!\begin{cases}
  \dfrac{ \frac{\sigma^2_{\rm n}}{M}e^{-\frac{\frac{\sigma^2_{\rm n}}{M} \gamma}{E\{\alpha\}(1-2\sigma^2_{\rm w}\gamma)}}}{E\{\alpha\} (1-2\sigma^2_{\rm w}\gamma)^2},    & 0\!\leqslant \!\gamma\!<\!\frac{1}{2\sigma^2_{\rm w}}, \\
   0,   & \text{otherwise}.
\end{cases}
  \label{Eqn:Eqn45}
\end{equation}

With the aid of \eqref{Eqn:Eqn45}, the BER expression of an uncoded QAM-based OFDM over the multipath Rayleigh fading channel is expressed by
\begin{equation}
  p_{\rm b}({\rm E})=\int\limits^{+\infty}_{0}{p_{\rm b}\left({\rm E}| \gamma \right) p_{\gamma}(\gamma)  d\gamma},
  \label{Eqn:Eqn46}
\end{equation}
where the conditional BER $p_{\rm b}\left({\rm E}| \gamma \right)$ of QAM is given by \cite{Ref30}
\begin{align}
&  p_{\rm b}\left({\rm E}| \gamma \right)=\frac{2}{\sqrt{J}\log_2\sqrt{J}} \sum\limits^{\log_2\sqrt{J}}_{u_1=1} \sum\limits^{(1-2^{-u_1})\sqrt{J}-1}_{u_2=0}  (-1)^{ \lfloor \frac{u_2 2^{u_1\!-\!1}}{\sqrt{J}}  \rfloor}   \nonumber \\
& \cdot \! \left( 2^{u_1\!-\!1}\!-\!\left\lfloor\! \frac{u_2 2^{u_1-1}}{\sqrt{J}}\! +\! \frac{1}{2}  \!\right\rfloor \right)  Q\!\!\left(\!\! (2u_2\!+\!1) \sqrt{\frac{3\log_2J}{J-1}} \gamma_{\rm b} \!\!\right) \!   ,
  \label{Eqn:Eqn47}
\end{align}
where $J$ is the size of the square QAM constellation, $\gamma_{\rm b}=\gamma/\log_2J$ denotes the SNR per bit, and $\lfloor x \rfloor$ indicates the largest integer to $x$.

Then, according to the proof in Appendix A, the asymptotic BER expression defined in \eqref{Eqn:Eqn46} is derived as
\begin{align}
&  p_{\rm b}({\rm E})=
  \frac{2}{\sqrt{J}\log_2\sqrt{J}} \sum\limits^{\log_2\sqrt{J}}_{u_1=1} \sum\limits^{(1-2^{-u_1})\sqrt{J}-1}_{u_2=0} (-1)^{\lfloor \frac{u_2 2^{u_1-1}}{\sqrt{J}} \rfloor}  \nonumber \\
& \cdot   \left(\! 2^{u_1-1}\!-\!\left \lfloor \frac{u_2 2^{u_1-1}}{\sqrt{J}} \!+\! \frac{1}{2}  \right \rfloor \! \right) \!\! \Bigg( \! \frac{1}{2}\!-\!\frac{1}{\sqrt{\pi}}\sum\limits^{+\infty}_{v_1=0} \sum\limits^{+\infty}_{v_2=0} \frac{(-1)^{v_1+v_2} }{v_1! (2v_1\!+\!1)}   \nonumber \\
& \cdot \left(\frac{3(2u_2+1)^2}{2(J-1)}\right)^{v_1+\frac{1}{2}}   \frac{\left(\frac{\sigma^2_{\rm n}}{M} \right)^{v_2+1}}{v_2!(E\{\alpha\})^{v_2+1}} \frac{(\sigma^{-})^{v_1+v_2+\frac{3}{2}}}{v_1+v_2+\frac{3}{2}}   \nonumber \\
& \cdot {}_2F_1(v_2+2,v_1+v_2+\frac{3}{2};v_1+v_2+\frac{5}{2};2\sigma^2_{\rm w}\sigma^{-}) \Bigg),
  \label{Eqn:Eqn48}
\end{align}
where ${}_2F_1(\cdot,\cdot;\cdot;\cdot)$ is the hypergeometric function and $\sigma^{-}=\lim\limits_{\epsilon>0 \; {\rm and} \; \epsilon\rightarrow 0}\{ \frac{1}{2\sigma^2_{\rm w}}- \epsilon \}$.
It can be inferred in \eqref{Eqn:Eqn48} that as the power of the smooth signal $2 \sigma^2_{\rm w}$ also increases, the BER $p_{\rm b}({\rm E})$ increases.

Finally, the average BER can be expressed as
\begin{equation}
p({\rm E})=\frac{1}{K}\sum\limits_{k_r\in \mathcal{K}}p_{{\rm b},k_r}({\rm E}),
  \label{Eqn:Eqn49}
\end{equation}
where $p_{{\rm b},k_r}({\rm E})$ is $p_{{\rm b}}({\rm E})$ in \eqref{Eqn:Eqn48} with the subcarrier index $k_r$.

\subsection{Error Performance Analysis with Imperfect Synchronization}

Without loss of generality, let us assume that the channel delays follow $0=\tau_{1}<\tau_{2}<\cdots<\tau_{\tilde{L}}=\tau_{\rm max}$.
In addition to the imperfect synchronization, both STO $\delta_{1}$ and CFO $\delta_{2}$ are considered at the receiver. 
Then, based on \cite{Ref33} and \eqref{Eqn:Eqn35}, the $i$th received signal $\bar{r}_i(t)$ can be expressed as
\begin{align}   \label{Eqn:Eqn50}
  \bar{r}_i(t) &= \sum\limits^{\tilde{L}}_{\tilde{l}=1}{h_{i,\tilde{l}}\bar{y}_i(t-\tau_{\tilde{l}}+\delta_{1}) e^{j2\pi\frac{\delta_{2}}{T_{\rm s}}(t-\tau_{\tilde{l}}+\delta_{1})} + n_i(t+\delta_{1}) }   \nonumber \\
  & = z_i(t+\delta_{1}) +u_i(t+\delta_{1}) + n_i(t+\delta_{1})
\end{align}
for $-T_{\rm cp} \leq t < T_{\rm s}+\tau_{\tilde{L}}$ , where
\begin{equation}   \label{Eqn:Eqn51}
z_i(t) = \sum\limits^{\tilde{L}}_{\tilde{l}=1} h_{i,\tilde{l}} y_i(t-\tau_{\tilde{l}}) e^{j2\pi\frac{\delta_{2}}{T_{\rm s}}(t-\tau_{\tilde{l}})},
\end{equation}
\begin{equation}   \label{Eqn:Eqn52}
u_i(t) = \sum\limits^{\tilde{L}}_{\tilde{l}=1} h_{i,\tilde{l}} w_i(t-\tau_{\tilde{l}}) e^{j2\pi\frac{\delta_{2}}{T_{\rm s}}(t-\tau_{\tilde{l}})}.
\end{equation}

\begin{figure}[tphb]
\centering
\includegraphics[width=3.3in]{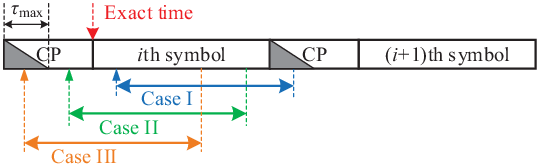}
\DeclareGraphicsExtensions.
\caption{Three cases of the imperfect time synchronization subject to STO.}
\label{Fig:Fig6}
\end{figure}

As shown in Fig. \ref{Fig:Fig6}, depending on the location of the estimated starting time of the OFDM symbol, three cases of the imperfect synchronization will be discussed below.
We first assume that the energy of the channel response is normalized, that is $\sum^{\tilde{L}}_{\tilde{l}=1} E\{ | h_{i,\tilde{l}} |^2\}=1$, and the data $x_{i,k_r}$ are i.i.d. random variables with zero mean and unit variance.
Meanwhile, we assume that the end of the channel-delayed smooth signal is after the estimated starting time, and before the exact end of the current OFDM symbol.

\subsubsection{Case I}
Since the estimated starting time of the OFDM symbol is after the exact time, as shown in Fig. \ref{Fig:Fig6}, after removing the CP, the received signal $\bar{r}_{i}(t)$ consists of a part of the current OFDM symbol $z_i(t)$ followed by a part of the next one $z_{i+1}(t)$, which is expressed as
\begin{equation}  \label{Eqn:Eqn53}
\bar{r}_{i}(t) \!=\!\begin{cases}
   z_i(t\!+\!\delta_{1}) \!+\! u_i(t\!+\!\delta_{1}) \!+\! n_i(t\!+\!\delta_{1}),   & 0 \!\leqslant\! t \!<\! T_{\rm s} \!-\! \delta_{1}, \\
  z_i(t\!+\!\delta_{1}) \!+\!  z_{i+1}(t\!+\!\delta_{1}\!-\!T)  \\
   \!+\! u_{i+1}(t\!+\!\delta_{1}\!-\!T) \!+\! n_{i+1}(t\!+\!\delta_{1}\!-\!T),   & T_{\rm s} \!-\! \delta_{1} \!\leqslant\! t \!<\! T_{\rm s}.
\end{cases} 
\end{equation}
To show the interference attributed to STO, CFO, and the smooth signal, Eq. \eqref{Eqn:Eqn53} can be rewritten as
\begin{equation}  \label{Eqn:Eqn54}
\bar{r}_{i}(t)= \hat{z}_i(t+\delta_1)+ \bar{z}_i(t), \;  0 \leqslant t <T_{\rm s},
\end{equation}
where 
\begin{equation}   \label{Eqn:Eqn82}
\hat{z}_i(t) = \sum\limits^{\tilde{L}}_{\tilde{l}=1} h_{i,\tilde{l}} y_i((t)_{T_{\rm s}}-\tau_{\tilde{l}}) e^{j2\pi\frac{\delta_{2}}{T_{\rm s}}(t-\tau_{\tilde{l}})},
\end{equation}
$(\cdot)_{T_{\rm s}}$ denotes the modulo of $T_{\rm s}$, for $0 \leqslant  t < T_{\rm s} - \delta_{1}$, 
\begin{equation}  \label{Eqn:Eqn55}
\bar{z}_i(t)=u_i(t+\delta_{1}) + n_i(t+\delta_{1}),
\end{equation}  
and for $T_{\rm s} - \delta_{1} \leqslant t < T_{\rm s}$, 
\begin{align}  \label{Eqn:Eqn56}
 \bar{z}_i(t)=
 &z_{i}(t + \delta_{1} )- \hat{z}_i(t+\delta_1) + z_{i+1}(t+\delta_{1} -T)  \nonumber \\
   & +u_{i+1}(t + \delta_{1} - T) +n_{i+1}(t+\delta_{1} -T),
\end{align}
where $z_{i}(t + \delta_{1} )- \hat{z}_i(t+\delta_1)$ and $z_{i+1}(t+\delta_{1} -T)$ represent the intra- and inter-symbol interferences, respectively. 

Based on the derivation in Appendix B, the average SINR of $\bar{r}_i(t)$ in \eqref{Eqn:Eqn54} can be expressed as 
\begin{align}  \label{Eqn:Eqn57}
\gamma_{\rm I}
&=\frac{\frac{1}{T_{\rm s}} \int\limits^{T_{\rm s}}_{0} E\left\{\left| \hat{z}_{i}(t+\delta_1)\right|^2\right\}dt}{\frac{1}{T_{\rm s}} \int\limits^{T_{\rm s}}_{0} E\left\{\left| \bar{z}_{i}(t)\right|^2\right\}dt}   \nonumber \\
&=\frac{K T_{\rm s}}{ \int\limits^{T_{\rm s}-\delta_1}_{0} E\left\{\left| \bar{z}_{i}(t)\right|^2\right\}dt +  \int\limits^{T_{\rm s}}_{T_{\rm s}-\delta_1} E\left\{\left| \bar{z}_{i}(t)\right|^2\right\}dt}  \nonumber \\
&=\frac{K T_{\rm s}}{ I_{1} +  \sigma^2_{\rm n} T_{\rm s} },
\end{align}
where ${\rm Re}\{\cdot\}$ represents the real part of a complex value and $I_{1}$ indicates the interference energy including STO, CFO, and the smooth signal, which is given by
\begin{align}  \label{Eqn:Eqn58}
I_{1}
&=  2 K \! \sum\limits^{L_{1}}_{\tilde{l}=1}  \! E\{|h_{i,\tilde{l}}|^2\} \! \left(\delta_1 \!-\! \tau_{\tilde{l}} \right) 
\!+\! \sum\limits^{N}_{n=0} \sum\limits^{N}_{\bar{n}=0} \! \sum\limits^{K-1}_{r=0} \! (a_{nr}a^{*}_{\bar{n}r}\!+\!b_{nr}b^{*}_{\bar{n}r})   \nonumber \\
& \cdot \Bigg(  \sum\limits^{\tilde{L}}_{\tilde{l}=1} \! E\{|h_{i,\tilde{l}}|^2\} \int\limits^{T_{\rm s}-\delta_1}_{0} \!\!\! \tilde{f}_n(t-\tau_{\tilde{l}}+\delta_1) \tilde{f}^{*}_{\bar{n}}(t-\tau_{\tilde{l}}+\delta_1) dt  \nonumber \\
& + \sum\limits^{L_{1}}_{\tilde{l}=1}  E\{|h_{i,\tilde{l}}|^2\}  \!\!\!\!\!\! \int\limits^{T_{\rm s}}_{T_{\rm s}\!-\! \delta_1 \!+\! \tau_{\tilde{l}}} \!\!\!\!\!\!\!\! \tilde{f}_n(t\!+\!\delta_1\!-\!T\!-\!\tau_{\tilde{l}}) \tilde{f}^{*}_{\bar{n}}(t\!+\!\delta_1\!-\!T\!-\!\tau_{\tilde{l}}) dt  \! \Bigg) \nonumber \\
&-2 \sum\limits^{L_{1}}_{\tilde{l}=1}  \sum\limits^{K-1}_{r=0} \! E\{|h_{i,\tilde{l}}|^2\}   {\rm Re}\Bigg\{ \! \sum\limits^{N}_{n=0}  \! \left( a_{nr} e^{-j 2\pi \frac{\delta_2 T + k_r T_{\rm cp}}{T_{\rm s}}} \!+\! b_{nr} \right)   \nonumber \\
&\quad \cdot  \!\!\!\!\!  \int\limits^{T_{\rm s}}_{T_{\rm s}-\delta_1+\tau_{\tilde{l}}}  \!\!\!\!\!\!\!\!  e^{-j2\pi\frac{k_r}{T_{\rm s}}(t\!+\!\delta_1\!-\!T\!-\!\tau_{\tilde{l}})} \tilde{f}_n(t\!+\!\delta_1\!-\!T\!-\!\tau_{\tilde{l}}) dt  \Bigg\},
\end{align}
where $L_{1}$ is the number of channel paths satisfying $\tau_{\tilde{l}} \leqslant \delta_{1}$.


\subsubsection{Case II}

As depicted in Fig. \ref{Fig:Fig6}, since the estimated starting time of the OFDM symbol is before the  exact time and after the end of the channel-delayed previous OFDM symbol $\bar{r}_{i-1}(t)$, there is no ISI in the current OFDM symbol $\bar{r}_i(t)$.
Then, for $0 \leqslant t <T_{\rm s}$, the $i$th received signal $\bar{r}_i(t)$ can be expressed as
\begin{equation}  \label{Eqn:Eqn59}
\bar{r}_{i}(t) =  z_i(t-\delta_{1}) + u_i(t-\delta_{1}) + n_i(t-\delta_{1}).
\end{equation}

Thus, according to \eqref{Eqn:Eqn10}, \eqref{Eqn:Eqn12}, \eqref{Eqn:Eqn51}, \eqref{Eqn:Eqn52}, and \eqref{Eqn:Eqn59}, the average SINR is expressed as
\begin{align}  \label{Eqn:Eqn60}
\gamma_{\rm II}
&=\frac{ \int\limits^{T_{\rm s}}_{0} E\left\{\left| z_{i}(t-\delta_1)\right|^2\right\}dt}{ \int\limits^{T_{\rm s}}_{0} E\left\{\left| u_{i}(t-\delta_1)\right|^2\right\}dt + \int\limits^{T_{\rm s}}_{0} E\left\{\left| n_{i}(t-\delta_1)\right|^2\right\}dt}   \nonumber \\
&= \frac{K T_{\rm s}}{I_{2}+\sigma^2_{\rm n} T_{\rm s}},
\end{align}
where the interference energy $I_2$ due to the STO and the smooth signal is given by
\begin{align}\label{Eqn:Eqn61}
I_{2} =
&\sum\limits^{\tilde{L}}_{\tilde{l}=1}  \sum\limits^{N}_{n=0} \sum\limits^{N}_{\bar{n}=0} \sum\limits^{K-1}_{r=0} \! (a_{nr}a^{*}_{\bar{n}r}\!+\!b_{nr}b^{*}_{\bar{n}r})  \nonumber\\
 & \cdot  E\{|h_{i,\tilde{l}}|^2\} \!\!\!\!\! \int\limits^{T_{\rm s}-\delta_1}_{0} \!\!\!\!\! \tilde{f}_n(t-\tau_{\tilde{l}}-\delta_1) \tilde{f}^{*}_{\bar{n}}(t-\tau_{\tilde{l}}-\delta_1) dt .
\end{align}

\subsubsection{Case III}
In this case, as shown in Fig. \ref{Fig:Fig6}, since the estimated starting time of the $i$th OFDM symbol is prior to the end of the channel-delayed $(i-1)$th OFDM symbol, the $i$th received signal $\bar{r}_i(t)$ is composed of a part of the current symbol $z_i(t)$ and a part of the previous symbol $z_{i-1}(t)$. 
Thus, for $0 \leqslant t <T_{\rm s}$, the $i$th received signal $\bar{r}_i(t)$ is expressed as
\begin{equation}  \label{Eqn:Eqn62}
\bar{r}_{i}(t) \!=\!\begin{cases}
   z_i(t\!-\!\delta_{1}) \!+\! u_i(t\!-\!\delta_{1}) \!+\! n_i(t\!-\!\delta_{1}) \\
  + z_{i-1}(t+T-\delta_1),   & 0 \!\leqslant\! t \!<\!\delta_1\!+\! \tau_{\tilde{L}} \!-\! T_{\rm cp} , \\
   z_i(t\!-\!\delta_{1}) \!+\! u_i(t\!-\!\delta_{1}) \!+\! n_i(t\!-\!\delta_{1}),   & \delta_1\!+\! \tau_{\tilde{L}} \!-\! T_{\rm cp} \!\leqslant\! t \!<\! T_{\rm s}.
\end{cases} 
\end{equation}
Similarly, to indicate the interference, Eq. \eqref{Eqn:Eqn62} is rewritten by
\begin{equation}  \label{Eqn:Eqn63}
\bar{r}_{i}(t)= \hat{z}_i(t-\delta_1)+ \tilde{z}_i(t) +n_i(t-\delta_1), \;  0 \leqslant t <T_{\rm s},
\end{equation}
where for $0 \!\leqslant\! t \!<\!\delta_1\!+\! \tau_{\tilde{L}} \!-\! T_{\rm cp} $, 
\begin{align}  \label{Eqn:Eqn64}
\tilde{z}_i(t) \!=\! z_i(t\!-\!\delta_1) \!-\! \hat{z}_i(t\!-\!\delta_1) \!+\! u_i(t\!-\!\delta_{1}) \!+\! z_{i-1}(t\!+\!T\!-\!\delta_1) ,
 \end{align}
and for $\delta_1\!+\! \tau_{\tilde{L}} \!-\! T_{\rm cp} \!\leqslant\! t \!<\! T_{\rm s}$,  
\begin{equation}  \label{Eqn:Eqn65}
\tilde{z}_i(t)= u_i(t\!-\!\delta_{1}),
 \end{equation}
where $z_i(t-\delta_1) - \hat{z}_i(t-\delta_1)$ and $z_{i-1}(t+T-\delta_1)$ are the intra- and inter-symbol interferences, respectively.

Via the energy analysis of the interference $\tilde{z}_i(t)$ in Appendix C, the average SINR is expressed as 
\begin{align}  \label{Eqn:Eqn66}
\gamma_{\rm III}
&= \frac{ \int\limits^{T_{\rm s}}_{0} E\{| \hat{z}_i(t-\delta_1) |^2\}dt}{\int\limits^{\delta_1\!+\! \tau_{\tilde{L}} \!-\! T_{\rm cp}}_{0} \! \!\! \! \! \! \!\! \!  E\{| \tilde{z}_i(t) |^2\}dt + \! \! \!\! \! \! \! \!\int\limits^{T_{\rm s}}_{\delta_1\!+\! \tau_{\tilde{L}} \!-\! T_{\rm cp}} \! \! \! \!\! \! \! \! \! E\{| \tilde{z}_i(t) |^2\}dt + \sigma^2_{\rm n}T_{\rm s}}  \nonumber \\
&= \frac{ K T_{\rm s} }{I_3 + \sigma^2_{\rm n}T_{\rm s}},
 \end{align}
where the interference energy $I_3$ is expressed as
\begin{align}  \label{Eqn:Eqn67}
I_{3}
=&  2 K \!\!\!\!\! \sum\limits^{\tilde{L}}_{\tilde{l}=\tilde{L}-L_{2}+1} \!\!\!\!\!\!\!  E\{|h_{i,\tilde{l}}|^2\}  ( \tau_{\tilde{l}}\!-\!T_{\rm cp}+\delta_1)    \nonumber \\
&+ \sum\limits^{N}_{n=0} \sum\limits^{N}_{\bar{n}=0} \sum\limits^{K-1}_{r=0} \! (a_{nr}a^{*}_{\bar{n}r}\!+\!b_{nr}b^{*}_{\bar{n}r})   \nonumber \\
&\cdot \Bigg( \sum\limits^{\tilde{L}}_{\tilde{l}=\tilde{L}-L_{2}+1}  \!\!\!\!\!\!\!\! E\{|h_{i,\tilde{l}}|^2\}   \!\!\!\!  \int\limits^{\delta_1\!+\! \tau_{\tilde{L}} \!-\! T_{\rm cp}}_{\delta_1\!+\! \tau_{\tilde{l}} \!-\! T_{\rm cp}} \!\!\!\!\! \!\!\!\!\!  \tilde{f}_n(t\!-\!\delta_1\!-\!\tau_{\tilde{l}}) \tilde{f}^{*}_{\bar{n}}(t\!-\!\delta_1\!-\!\tau_{\tilde{l}}) dt    \nonumber \\
& + \sum\limits^{\tilde{L}-L_{2}}_{\tilde{l}=1}  \!\!\! E\{|h_{i,\tilde{l}}|^2\}   \!\!\!\!\!  \int\limits^{\delta_1\!+\! \tau_{\tilde{L}} \!-\! T_{\rm cp}}_{0} \!\!\!\!\! \!\!\!\!\!  \tilde{f}_n(t\!-\!\delta_1\!-\!\tau_{\tilde{l}}) \tilde{f}^{*}_{\bar{n}}(t\!-\!\delta_1\!-\!\tau_{\tilde{l}}) dt  \nonumber \\
&+  \sum\limits^{\tilde{L}}_{\tilde{l}=1}   E\{|h_{i,\tilde{l}}|^2\}  \int\limits^{T_{\rm s}}_{\delta_1\!+\! \tau_{\tilde{L}} \!-\! T_{\rm cp}} \!\!\!\!\! \!\!\!\!\!  \tilde{f}_n(t\!-\!\delta_1\!-\!\tau_{\tilde{l}}) \tilde{f}^{*}_{\bar{n}}(t\!-\!\delta_1\!-\!\tau_{\tilde{l}}) dt \Bigg),
\end{align}
where $L_{2}$ represents the number of channel paths satisfying $\tau_{\tilde{l}} \geqslant T_{\rm cp}-\delta_{1}$.

\begin{figure}[h]
\centering
\includegraphics[width=3.7in]{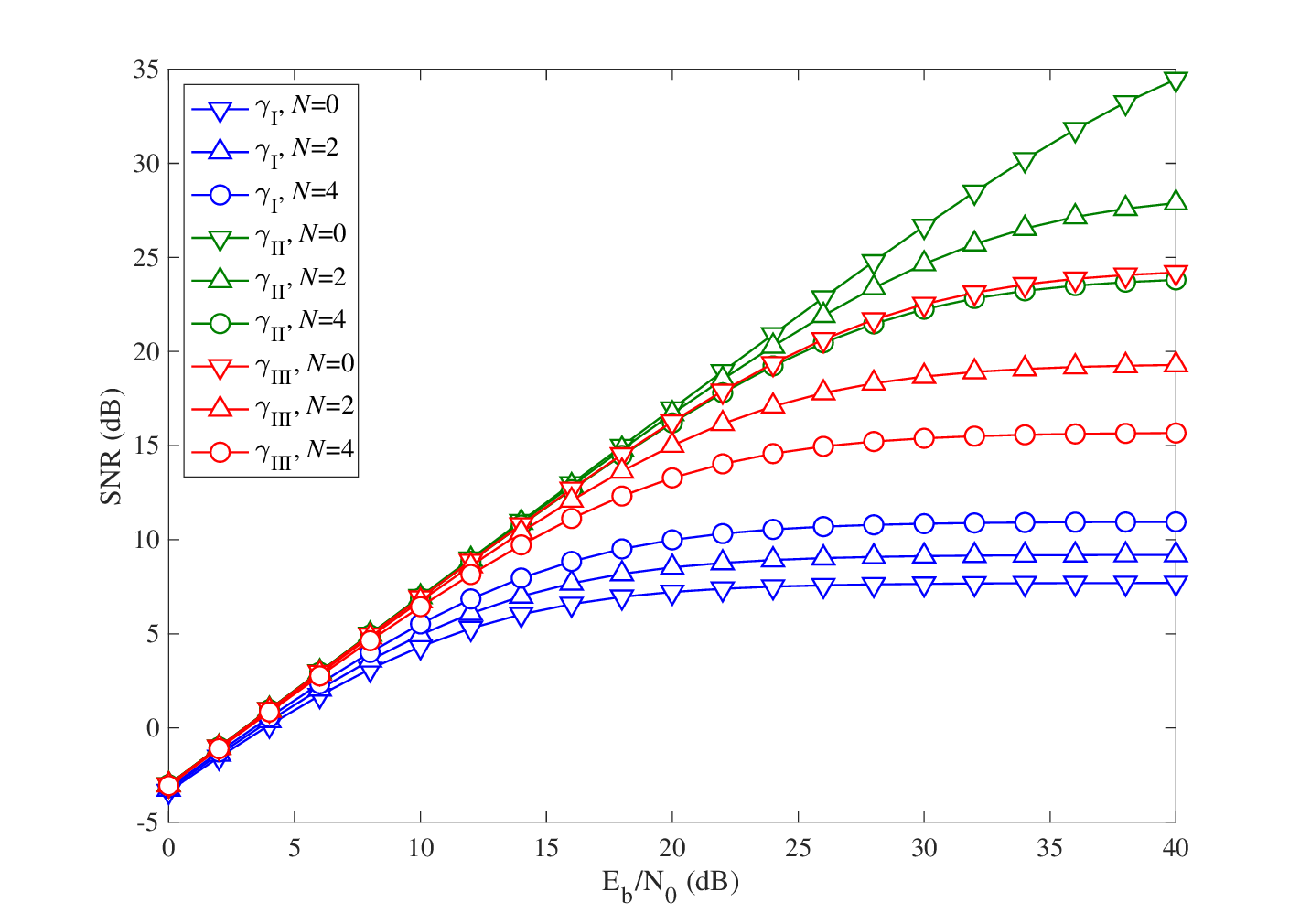}
\DeclareGraphicsExtensions.
\caption{Average SINR versus per-bit SNR in an OFDM system with imperfect synchronization exposed to Rayleigh fading (9-path EVA channel in \cite{Ref27}), when using 256 subcarriers and 16QAM. 
The value of CFO normalized by the subcarrier spacing is 0.074. 
The length of the smooth signal normalized by the sampling interval is $L=1000$. 
The values of STO normalized by the sampling interval are 30, 30, and 97 for Cases I, II, and III, respectively.  }
\label{Fig:Fig7}
\end{figure}

Fig. \ref{Fig:Fig7} plots the analyzed average SINR versus per-bit SNR of the above three cases. 
From the curves we observe that for a fixed value of per-bit SNR and the given values of $\delta_1$ and $\delta_2$, the increased $N$ will lead to a small decrease in SINR.
Furthermore, the high intra- and inter-symbol interferences associated with STO and CFO in Cases I and III cause a much lower $\gamma_{\rm I}$ and $\gamma_{\rm III}$ compared to $\gamma_{\rm II}$, due to no intra- and inter-symbol interferences in Case II. 

\begin{changebar}{\color{blue}

}\end{changebar}

\subsection{Analysis of ${\rm E_b/N_0}$}

According to \cite{Ref34}, the linear relationship in dB between per-bit SNR and the average SINR $\bar{\gamma}_{\rm SINR}$ over AWGN channel can be expressed as
\begin{equation}\label{Eqn:Eqn68}
{\rm E_b/N_0}= 10\log_{10}\bar{\gamma}_{\rm SINR} + 10\log_{10}(\bar{J}) -10\log_{10}(\log_{2}J),
\end{equation}
where $\bar{J}$ represents the time-domain oversampling factor. 

According to \eqref{Eqn:Eqn10} and \eqref{Eqn:Eqn35}, $\bar{\gamma}_{\rm SINR}$ can be expressed as
\begin{align}  \label{Eqn:Eqn69}
\bar{\gamma}_{\rm SINR}
&=  \frac{ \frac{1}{T} \int\limits^{T_{\rm s}}_{-T_{\rm cp}} \!\!\!\!  E\left\{\left| y_i(t) \right|^2  \right\} dt }{ \frac{1}{T} \int\limits^{T_{\rm s}}_{-T_{\rm cp}} \!\!\!\!  E\left\{\left| w_i(t) \right|^2  \right\} dt+ \frac{1}{T}\int\limits^{T_{\rm s}}_{-T_{\rm cp}}  \!\!\!\! E\left\{\left| n_i(t) \right|^2  \right\} dt } \nonumber \\
&= \frac{ K T }{ \int\limits^{T_{\rm s}}_{-T_{\rm cp}} \!\!\! E\left\{ \left|w_i(t) \right|^2  \right\}  dt + T \sigma^2_{\rm n} } .
\end{align}

Meanwhile, the energy of $w_i(t)$ in \eqref{Eqn:Eqn12} can be expressed as
\begin{align}\label{Eqn:Eqn70}
&\int\limits^{T_{\rm s}}_{-T_{\rm cp}} E\left\{ \left|w_i(t) \right|^2  \right\} dt    \nonumber \\
&\quad =\sum\limits^{N}_{n=0} \sum\limits^{N}_{\bar{n}=0} \sum\limits^{K-1}_{r=0}  \left(  a^*_{nr} a_{\bar{n}r}   +  b^*_{nr} b_{\bar{n}r}  \right) \!\! \int\limits^{T_{\rm s}}_{-T_{\rm cp}} \!\!\! \tilde{f}^*_n(t) \tilde{f}_{\bar{n}}(t) dt .
\end{align}

Thus, it follows from \eqref{Eqn:Eqn68} and \eqref{Eqn:Eqn70} that the relationship between ${\rm E_b/N_0}$ and $w_i(t)$ can be expressed as
\begin{align}\label{Eqn:Eqn71}
&{\rm E_b/N_0}= 10\log_{10} \frac{ K \bar{J} }{\log_{2}J} - 10\log_{10} \Bigg(\!  \sigma^2_{\rm n}  \nonumber \\
&\quad  + \sum\limits^{N}_{n=0} \sum\limits^{N}_{\bar{n}=0} \sum\limits^{K-1}_{r=0} \! \left(  a^*_{nr} a_{\bar{n}r}   \!+\!  b^*_{nr} b_{\bar{n}r}  \right) \!\!\! \int\limits^{T_{\rm s}}_{-T_{\rm cp}} \!\!\! \frac{\tilde{f}^*_n(t) \tilde{f}_{\bar{n}}(t)}{T} dt  \! \Bigg).
\end{align}

\begin{figure}[t]
\centering
\includegraphics[width=3.7in]{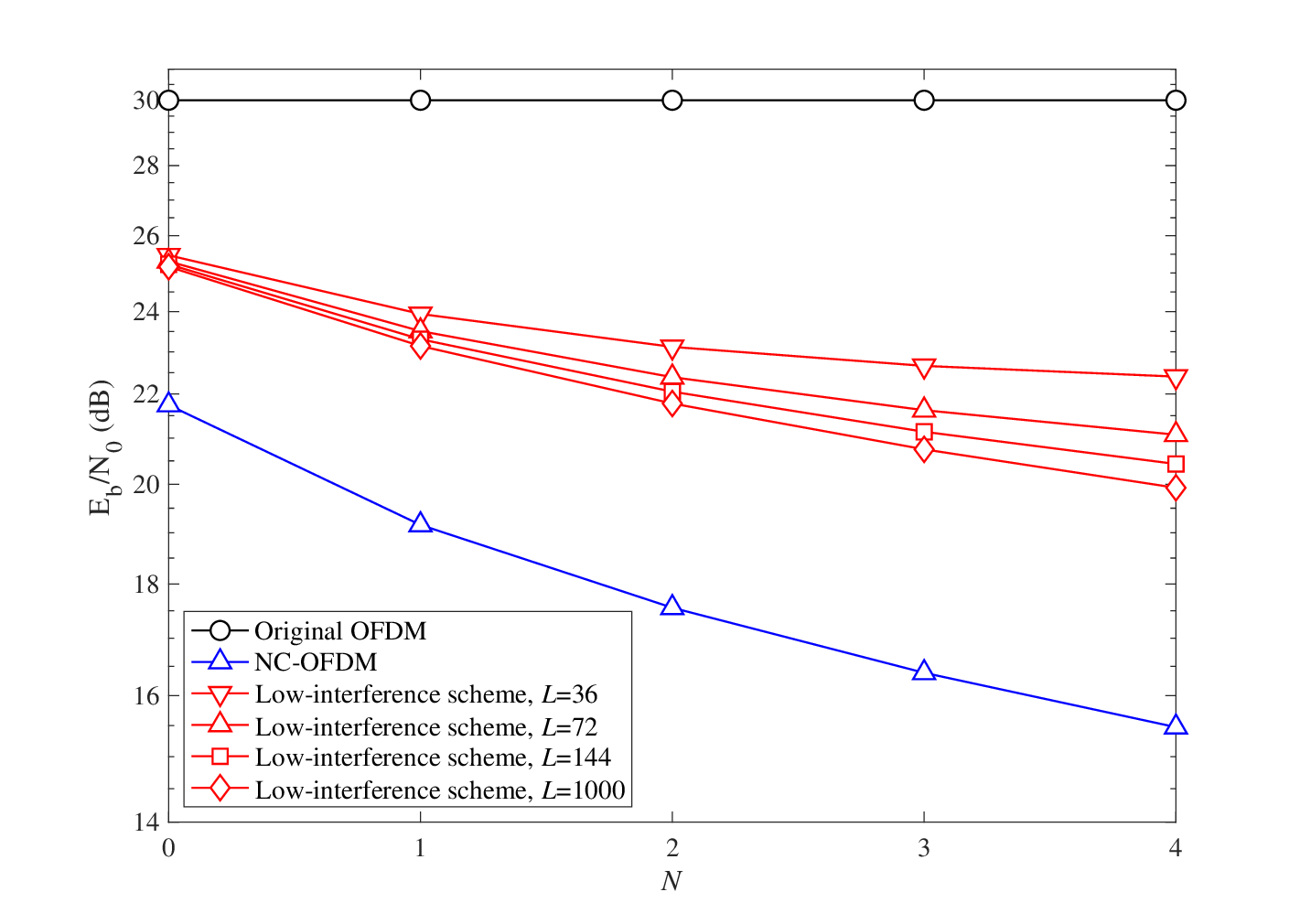}
\DeclareGraphicsExtensions.
\caption{${\rm E_b/N_0}$ versus the HDO $N$ according to \eqref{Eqn:Eqn71} for conventional NC-OFDM and low-interference NC-OFDM, when using 16QAM. 
The reference value of ${\rm E_b/N_0}$ in original OFDM is 30dB. 
The lengths of the smooth signal are $L=36, 72, 144, 1000$. The oversampling factor is $\bar{J}=8$.}
\label{Fig:Fig8}
\end{figure}

It can be shown in Fig. \ref{Fig:Fig8} that the analyzed ${\rm E_b/N_0}$ of \eqref{Eqn:Eqn71} is reduced by increasing $N$ and $L$, which simultaneously increases the power of $w_i(t)$.
Furthermore, the ${\rm E_b/N_0}$ of the low-interference scheme is significantly higher than that of the conventional NC-OFDM, and close to that of the original OFDM.
Corresponding to the analysis in Section IV-A, it is also implied that the low-interference scheme is capable of achieving a small error performance degradation as opposed to original OFDM.

\subsection{Error Variance Analysis of Time Synchronization}

In this subsection, we will focus our attention on the error performance of the STO estimation to obtain the effective solution of time synchronization for the low-interference scheme.
According to \cite{Ref33} and \cite{Ref37,Ref38,Ref39}, due to the cyclostationarity of the CP-OFDM signal, the correlation function of the received OFDM signal is important for time synchronization. 
However, since the smooth signal is just added into the front part $[-T_{\rm cp},-T_{\rm cp}+T_{L}]$ of each CP-OFDM symbol, it destroys the cyclostationarity between the CP in $[-T_{\rm cp},-T_{\rm cp}+T_{L}]$ and its corresponding copy in $[-T_{\rm cp}+T_{L}, T_{\rm s}]$.
Then, we first define the correlation function of the received OFDM signal $\bar{r}_i(t)$ as $R(t,\Delta)=\frac{1}{K} E\left\{ \bar{r}_i(t) \bar{r}^{*}_i(t+\Delta) \right\}$.
Thus, for $t \in [-T_{\rm cp},-T_{\rm cp}+T_{L}]$ and $t+\Delta \in [-T_{\rm cp}+T_{L}, T_{\rm s}]$, $R(t,\Delta)$ is expressed as
\begin{align}\label{Eqn:Eqn83}
R(t,\Delta)
 &= \frac{1}{K} e^{-j2\pi \frac{\delta_2}{T_{\rm s}}\Delta}  \sum\limits^{\tilde{L}}_{\tilde{l}=1} \sum\limits^{K-1}_{r=0} E\{ | h_{i,\tilde{l}} |^2\}   \Bigg( \!  e^{-j2\pi \frac{k_r}{T_{\rm s}}\Delta}  \nonumber \\
&  - \sum\limits^{N}_{n=0} b_{nr} \tilde{f}_n(t\!-\! \tau_{\tilde{l}} \!+\! \delta_1) e^{-j2\pi \frac{k_r}{T_{\rm s}}(t+\Delta-\tau_{\tilde{l}}+\delta_1)}  \! \Bigg)\!.
\end{align}

According to \cite{Ref37}, for the CP-OFDM signal, if no subcarrier weighting is employed, $R(t,\Delta)$ in \eqref{Eqn:Eqn83} includes the information of the time synchronization parameters for $\Delta=T_{\rm s}$ only.
In this case, under the assumption of $T_{L} \leqslant T_{\rm s}$, Eq. \eqref{Eqn:Eqn83} can be rewritten as
\begin{align}\label{Eqn:Eqn84}
R(t,T_{\rm s})= 
& e^{-j2\pi \delta_2} \Bigg(   1 - \frac{1}{K} \sum\limits^{\tilde{L}}_{\tilde{l}=1} \sum\limits^{K-1}_{r=0} E\{ | h_{i,\tilde{l}} |^2\}  \nonumber \\
& \cdot \sum\limits^{N}_{n=0} b_{nr} \tilde{f}_n(t-\tau_{\tilde{l}}+\delta_1) e^{-j2\pi \frac{k_r}{T_{\rm s}}(t-\tau_{\tilde{l}}+\delta_1)}   \Bigg).
\end{align}

\begin{figure}[h]
\centering
\includegraphics[width=3.7in]{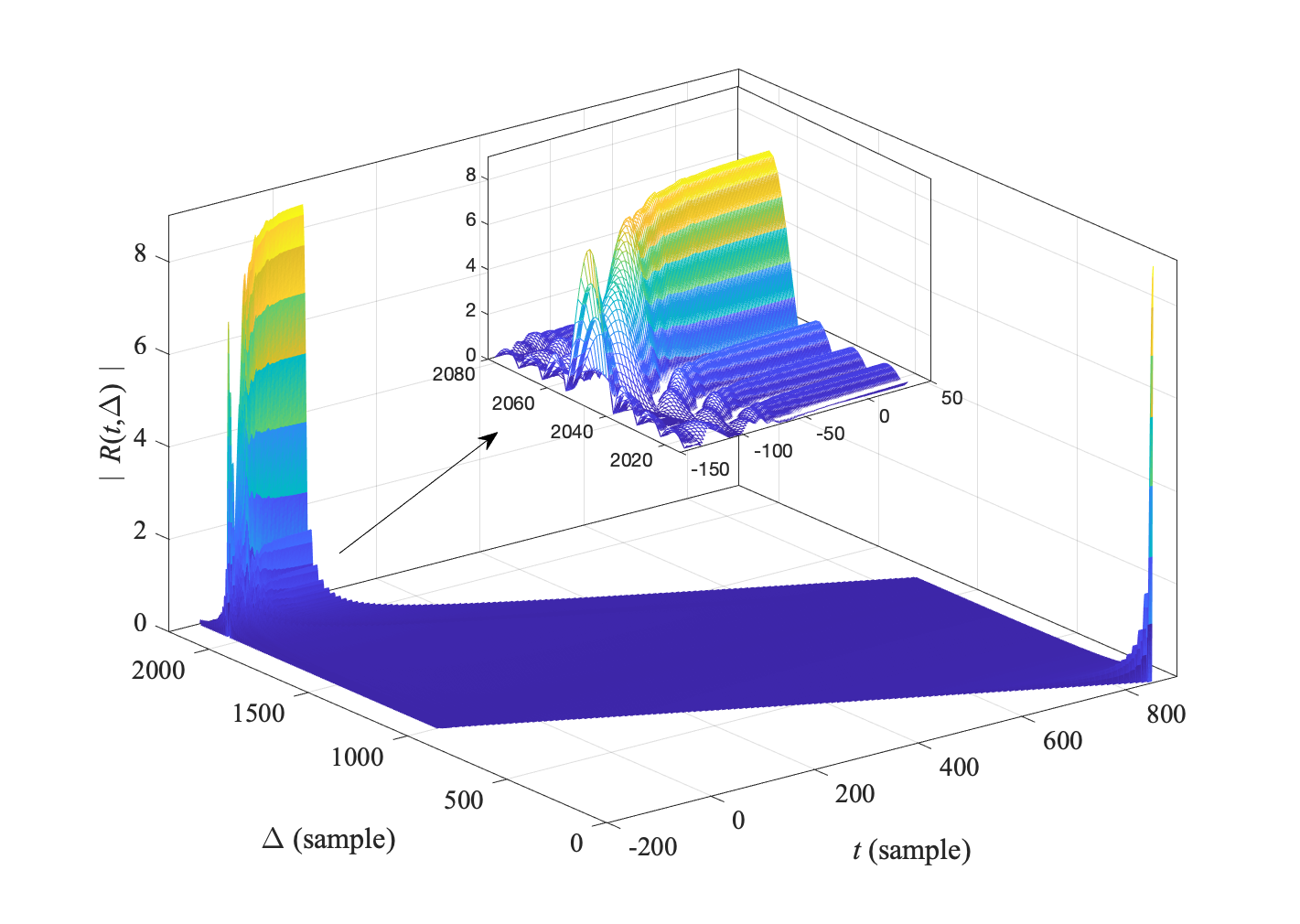}
\caption{$\left| R(t,T_{\rm s}) \right|$ versus $t$ and $\Delta$ over a multipath Rayleigh fading channel, assuming 256 subcarriers, $\delta_1=0$s, and $\delta_2=0$Hz. 
The length of the smooth signal is $L=1000$.
The value of HDO is $N=4$.}
\label{Fig:Fig14}
\end{figure}

\begin{figure}[h]
\centering
\includegraphics[width=3.7in]{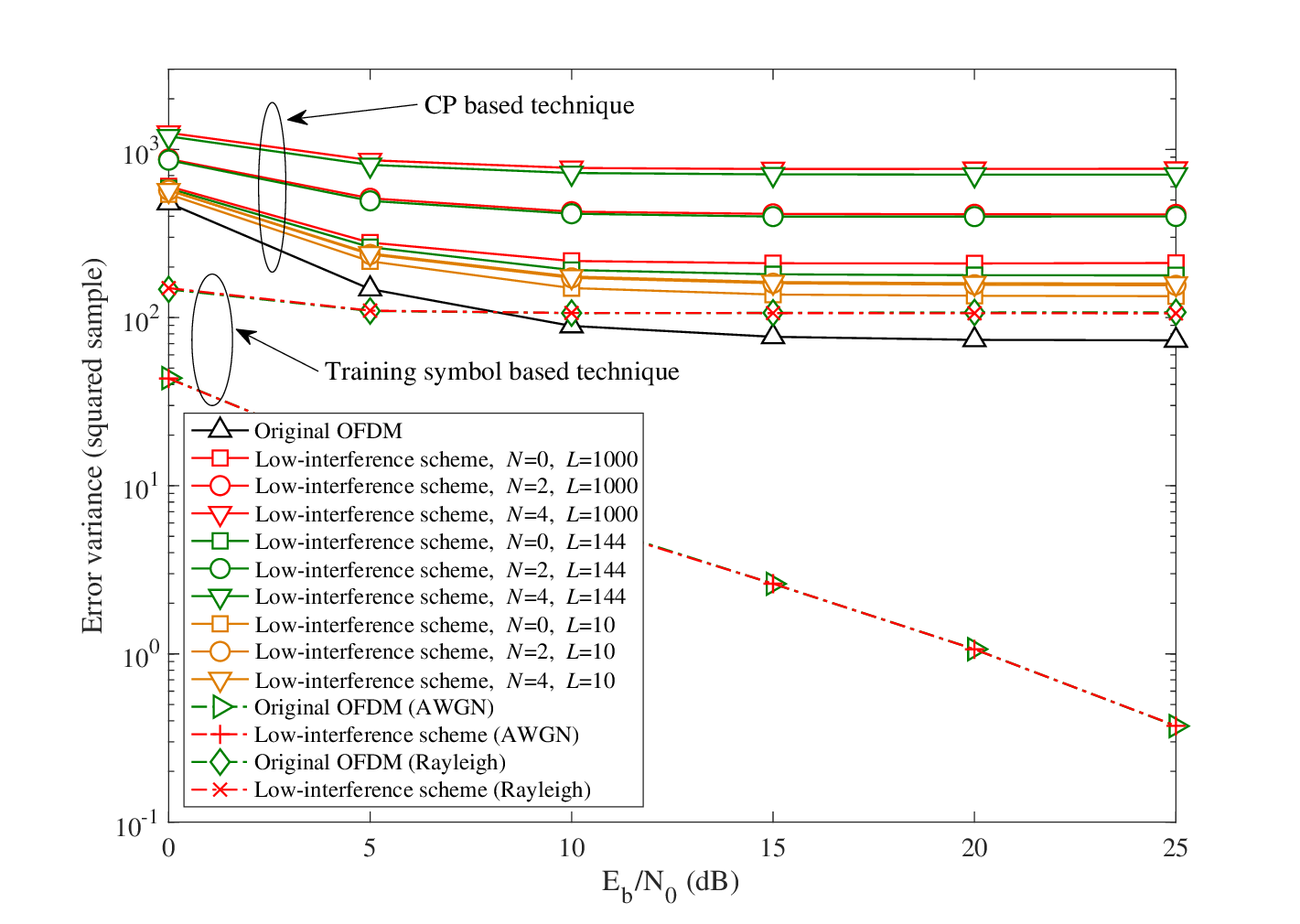}
\caption{Mean-squared error variance of the estimated time offset versus per-bit SNR of original OFDM and low-interference scheme, assuming 256 subcarriers, 16QAM, and $\delta_2=111.11$Hz. 
The value of STO normalized by the sampling interval is 30 for Case I.}
\label{Fig:Fig15}
\end{figure}

Fig. \ref{Fig:Fig14} verifies that the maximum value of $|R(t,\Delta)|$ is associated with $\Delta=T_{\rm s}$ for $t\in [-T_{\rm cp}, 0]$.
However, due to the smooth signal, $R(t,T_{\rm s})$ in \eqref{Eqn:Eqn84} and Fig. \ref{Fig:Fig14} imply that $\bar{r}_i(t)$ does not completely follow the cyclostationarity distribution.
Furthermore, as $t$ decreases from $0$ to $-T_{\rm cp}$, corresponding to the valley of $|R(t,T_{\rm s})|$ in Fig. \ref{Fig:Fig14}, the weakened correlation will cause an accuracy reduction for time synchronization. 
On the contrary, when increasing $t$ to 0, the reduced instantaneous power of the smooth signal increases the value of $|R(t,T_{\rm s})|$.
Consequently, decreasing the value of $N$ or $L$ is capable of improving the performance of time synchronization.
As observed in Fig. \ref{Fig:Fig15}, upon using CP based STO estimation technique \cite{Ref33,Ref38} over a Rayleigh fading channel (EVA channel in \cite{Ref27}), when the value of $N$ or $L$ decreases, the mean-squared error variances of the low-interference scheme are effectively reduced.
However, since most power of the smooth signal is concentrated in the front part of each OFDM symbol, if the value of $N$ or $L$ is inappropriately reduced, the sidelobe suppression performance will be significantly degraded. 

In order to mitigate the effect of the smooth signal on time synchronization and maintain a compact spectrum, invoked by the small power ratio of the smooth signal in $[0, T_{\rm s}]$, the STO estimation using training (reference) symbol is employed \cite{Ref33,Ref39}.
Fig. \ref{Fig:Fig15} also presents the mean-squared error variance versus per-bit SNR of original OFDM and low-interference scheme for AWGN and Rayleigh fading channels, when using the training symbol based STO estimation technique.  
The length and HDO of the smooth signal are $L=1000$ and $N=4$, respectively.
For large values of $N$ and $L$, we notice that the error variances of the low-interference scheme are close to those of original OFDM.

\section{Numerical Results}

In this section, we present our performance results characterizing the low-interference \emph{N}-continuous OFDM system employing 16-QAM on \emph{K}=256 subcarriers, having a subcarrier spacing of $\Delta f=1/T_{\rm s}$=15KHz and transmitting the signal with a sampling interval $T_{\rm samp}=T_{\rm s}/2048$ and a CP duration of $T_{\rm cp}=144 T_{\rm samp}$.
For portraying the PSD performance, the Welch's averaged periodogram method \cite{Ref29} is employed under the setting of a 2048-sample Hanning window and 512-sample overlap. %
In order to show the BER performance in the multipath fading environment, we adopt the EVA channel model provided in \cite{Ref27} having the excess tap delay [0, 30, 150, 310, 370, 710, 1090, 1730, 2510] ns and the relative power [0, -1.5, -1.4, -3.6, -0.6, -9.1, -7, -12, -16.9] dB.

\begin{figure}[t]
\centering
\includegraphics[width=3.7in]{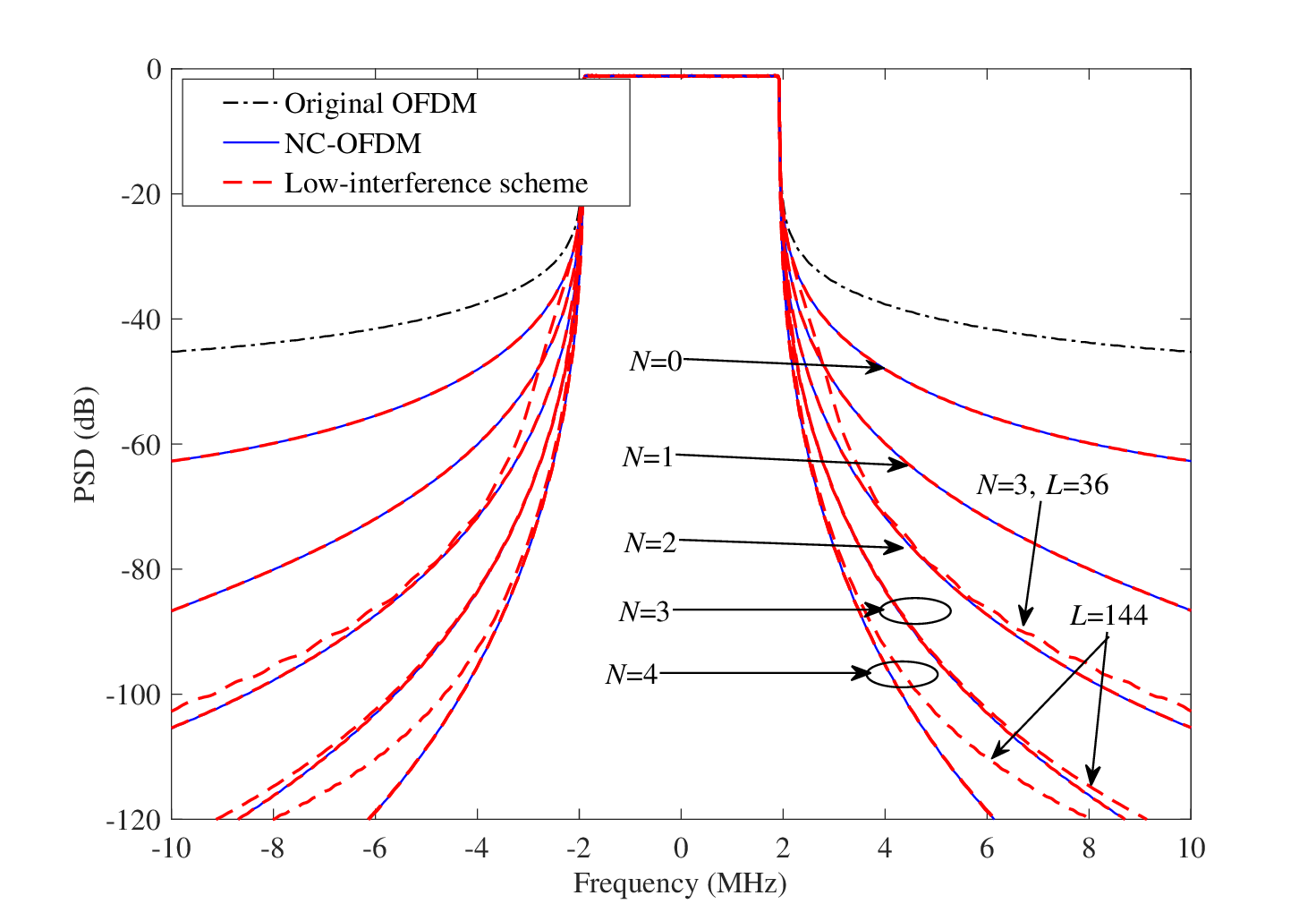}
\caption{PSDs of original OFDM, conventional NC-OFDM, and low-interference NC-OFDM with varying $N$ and $L$.}
\label{Fig:Fig9}
\end{figure}

Fig. \ref{Fig:Fig9} shows the impact of the HDO \emph{N} and the length of the smooth signal \emph{L} on the PSD of the low-interference scheme as opposed to conventional OFDM and NC-OFDM, where $N=0, 1, 2, 3, 4$ and $L=144, 500, 1000 $.
As can be observed from the results, the low-interference scheme can achieve sidelobe decaying close to conventional NC-OFDM.
When \emph{N} is large and \emph{L} is small, e.g. \emph{N}=3 and \emph{L}=36, the sidelobe decaying rate of the low-interference scheme significantly reduces as opposed to NC-OFDM.
As expected, with an increase in \emph{L}, such as from \emph{L}=36 to \emph{L}=144, a more steep OOB spectrum is attainable.

\begin{figure}[t]
\centering
\includegraphics[width=3.7in]{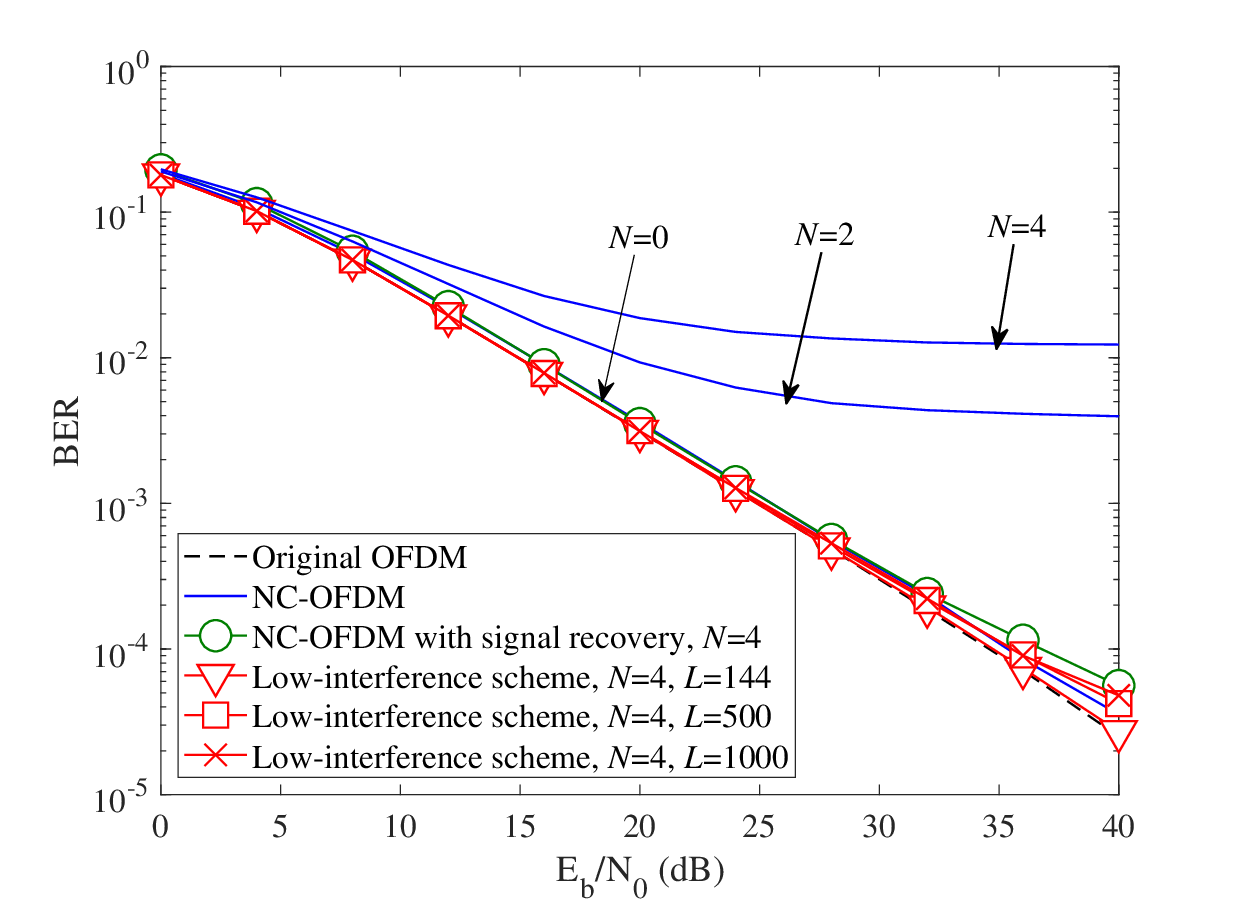}
\caption{BER versus per-bit SNR performance of original OFDM, conventional NC-OFDM, and low-interference NC-OFDM over a multipath Rayleigh fading channel, assuming perfect synchronization, perfect channel estimation, and various values of $N$ and $L$.}
\label{Fig:Fig10}
\end{figure}

Fig. \ref{Fig:Fig10} illustrates the average BER performance versus the per-bit SNR among original uncoded OFDM, conventional NC-OFDM, and the low-interference NC-OFDM scheme, using the zero-forcing (ZF) channel equalization over the multipath Rayleigh fading channel.
It confirms, not surprisingly, that the smooth signal has a slightly negative effect on the BER of the low-interference scheme in the multipath fading channel, corresponding to the analyses of the SINR and BER in Section IV.
For example, when \emph{N}=4, compared to OFDM, the BER performance of NC-OFDM is severely degraded.
On the contrary, the low-interference scheme is capable of achieving a considerable BER, even if the value of \emph{L} is increased to 1000.
Although the interference in NC-OFDM can be well eliminated by a signal recovery algorithm \cite{Ref13} with $I_{\rm R}=8$ iterations, compared to the low-interference scheme, extreme computational complexity is introduced at the receiver.

\begin{figure}[h]
\centering
\includegraphics[width=3.7in]{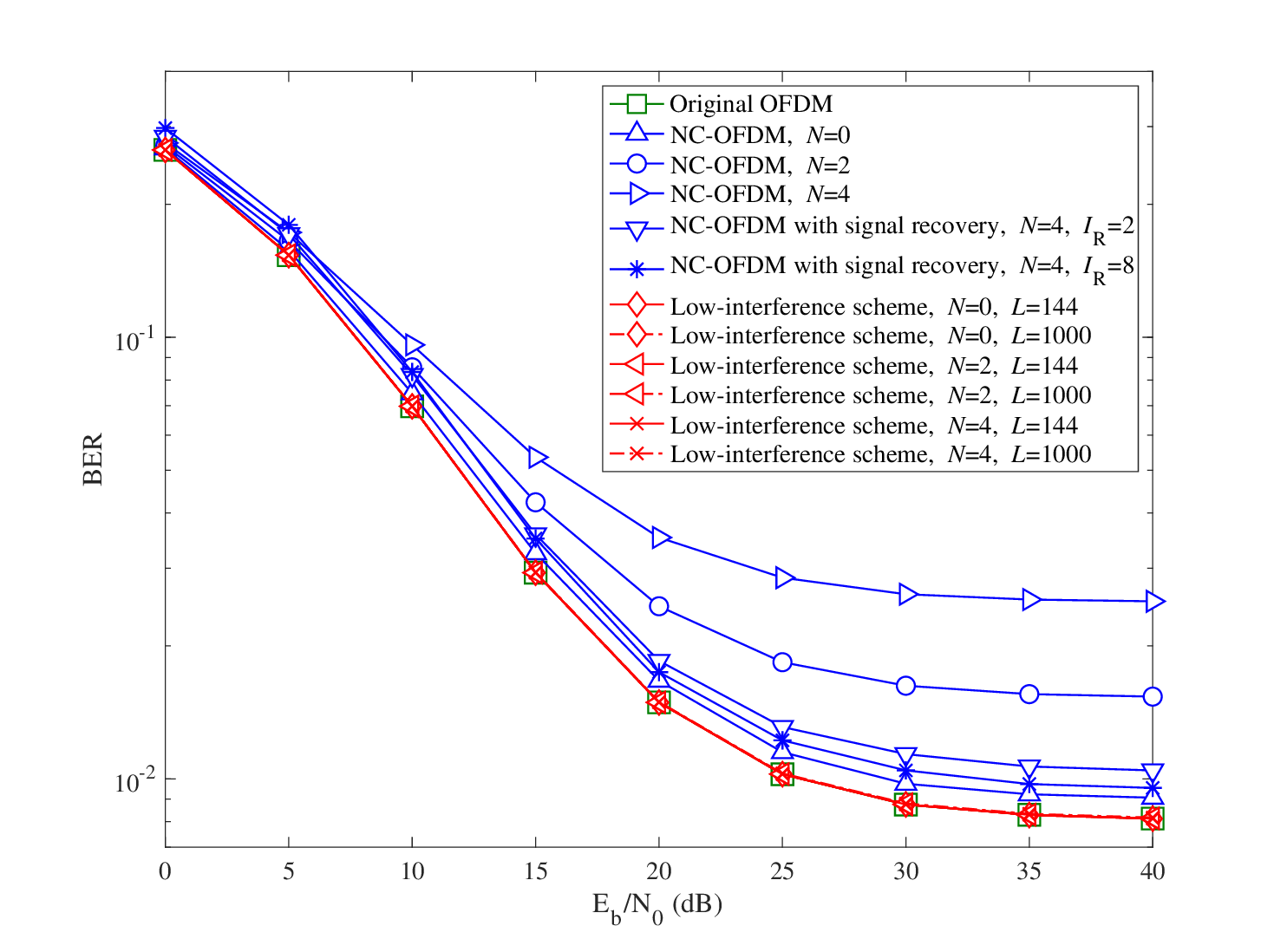}
\caption{BER versus per-bit SNR performance of original OFDM, conventional NC-OFDM, and low-interference NC-OFDM over a multipath Rayleigh fading channel, assuming $\delta_1=0.033 \mu$s, $\delta_2=111.11$Hz, and various values of $N$ and $L$.}
\label{Fig:Fig11}
\end{figure}

Furthermore, the above observation can be further augmented with the aid of Fig. \ref{Fig:Fig11}, where the BER versus per-bit SNR performance of original OFDM, conventional NC-OFDM, and low-interference NC-OFDM is compared over the multipath Rayleigh fading channel, using the parameters of $N=0,2,4$ and $L=144,1000$ for the NC-OFDM schemes, and $I_{\rm R}=2,8$ for the number of signal recovery algorithm at conventional NC-OFDM receiver.
The STO associated with Case I in Section IV-B is $\delta_1=0.033 \mu$s and the CFO is $\delta_2=111.11$Hz.
Meanwhile, the first one out of each two consecutive OFDM symbols is used as the reference signal for the least-squared (LS) channel estimation.
Observe from Fig. \ref{Fig:Fig11} that as expected, for a constant SNR per bit, at the OFDM receiver affected by STO and CFO, the low-interference scheme still outperforms conventional NC-OFDM, while having the BER curves close to original OFDM. 
The results also show that, with the increased $I_{\rm R}$, the error floor of the recovered signal at the conventional NC-OFDM receiver is a little lowered due to the STO and CFO. 

\begin{figure}[t]
\centering
\subfloat[ ]{\includegraphics[width=3.7in]{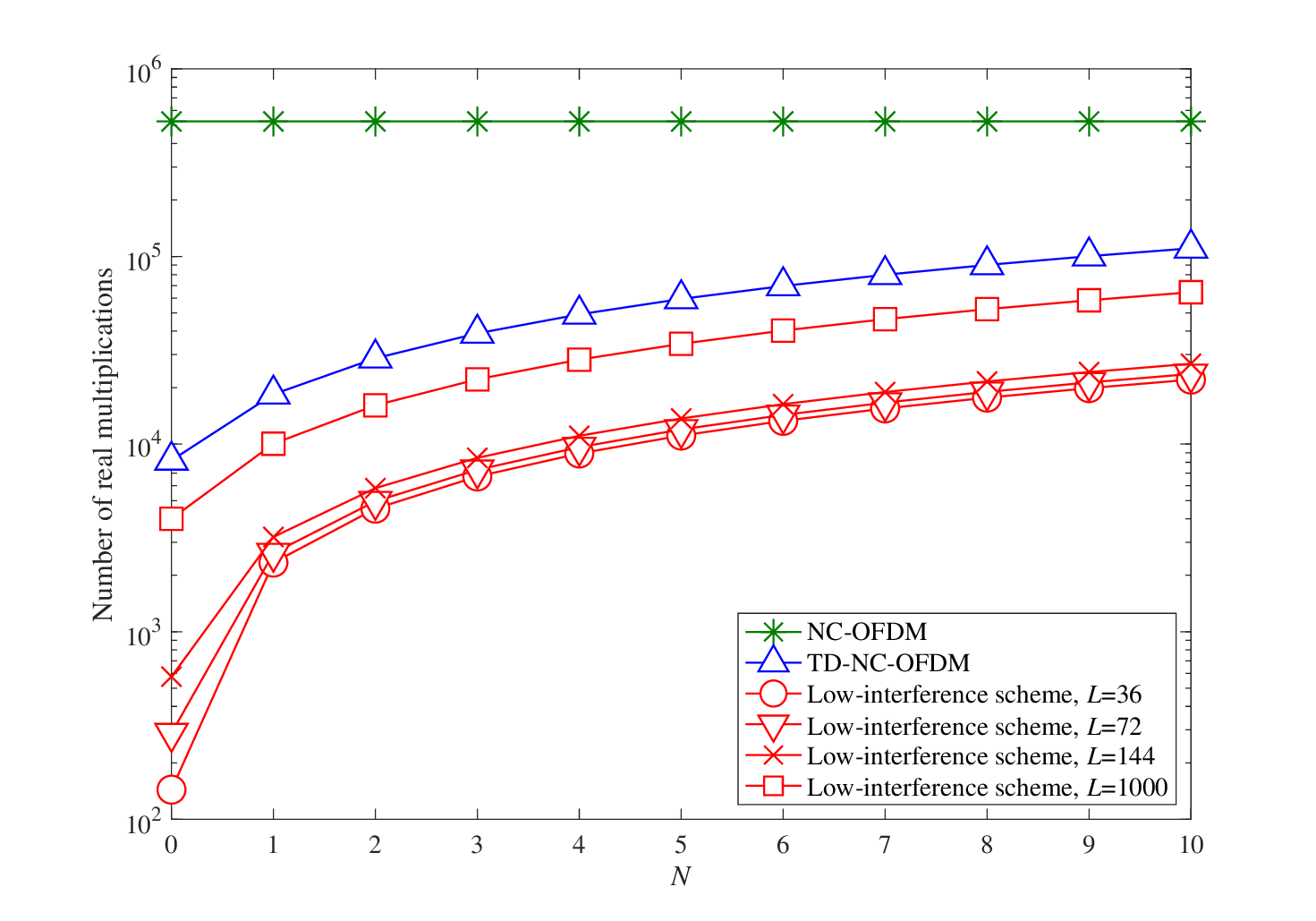}
\label{Fig:Fig12a}}
\hfil
\subfloat[ ]{\includegraphics[width=3.7in]{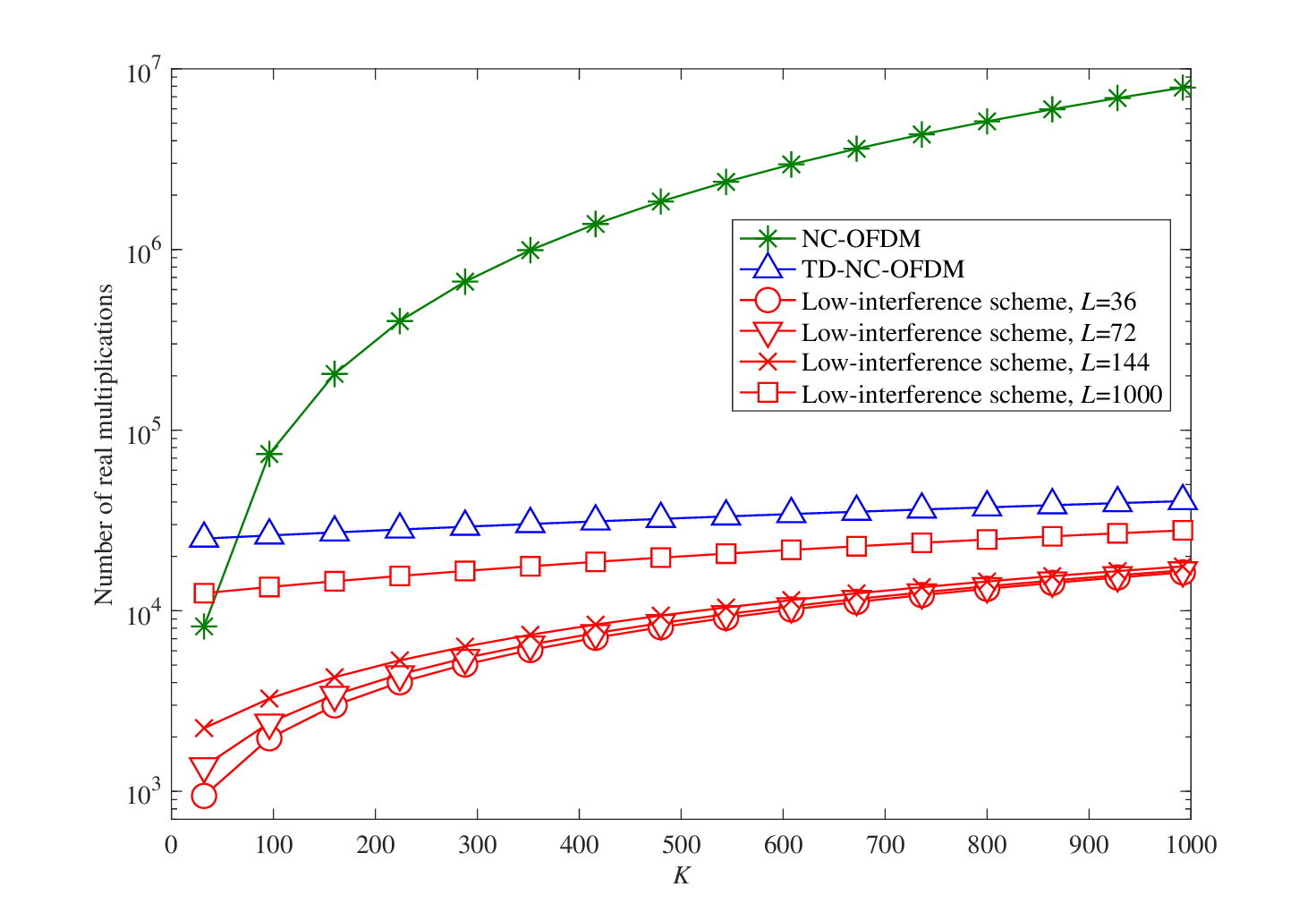}
\label{Fig:Fig12b}}
\caption{Complexity comparison of conventional NC-OFDM, TD-NC-OFDM, and low-interference NC-OFDM at the transmitter. (a) $K=256$, (b) $N=2$.}
 \label{Fig:Fig12}
\end{figure}

Additionally, the complexity of the low-interference NC-OFDM scheme is shown in Fig. \ref{Fig:Fig12}.
According to \cite{Ref24} and \cite{Ref25}, we plot the number of real multiplications of the low-interference NC-OFDM scheme, conventional NC-OFDM, and TD-NC-OFDM at the transmitter, using the parameters of $L=36,72,144, 1000$, a constant number of subcarriers $K=256$ for Fig. \ref{Fig:Fig12a}, and a fixed value of HDO $N=2$ for Fig. \ref{Fig:Fig12b}.
As observed in Fig. \ref{Fig:Fig12}, compared to conventional NC-OFDM and TD-NC-OFDM, the low-interference NC-OFDM scheme has a much higher computational efficiency, while it can maintain a compact spectrum close to conventional NC-OFDM and TD-NC-OFDM as shown in Fig. \ref{Fig:Fig9} and a small BER performance degradation as opposed to original OFDM as shown in Figs. \ref{Fig:Fig10} and \ref{Fig:Fig11}.
Furthermore, contrary to the conventional NC-OFDM receiver, the low-interference scheme is capable of avoiding extra signal recovery at the receiver, due to its slight BER performance degradation.
Meanwhile, the low-interference scheme is capable of attaining the same PAPR as original OFDM, and so are NC-OFDM and TD-NC-OFDM systems.

\section{Conclusion}

In this contribution, the PSD and error performances of the low-interference NC-OFDM \cite{Ref24, Ref25} with various HDOs and lengths of the smooth signal were thoroughly investigated and compared.
The PSD was first evaluated in the form of an asymptotic expression with the aid of the introduced smooth signal.
It was concluded from our analytical and simulated results that, when the HDO and the length of the smooth signal increase, significant improvement of sidelobe suppression will be observed.
However, as the length of the smooth signal reduces, the sidelobe suppression performance of the low-interference scheme might be significantly degraded, even if using a large HDO.
Then, assuming perfect synchronization, we provided an asymptotic expression for quantifying the BER of the low-interference scheme in the multipath fading channel by deriving a closed-form expression of the instantaneous SINR associated with the channel delay.
Furthermore, the average SINR performance of the low-interference scheme was evaluated over the multipath fading channel, when both STO and CFO were considered at the receiver.
We also have studied the effects of the length $L$ and the number of basis signals $N$ of the smooth signal on the per-bit SNR performance of the low-interference scheme.
It revealed that the increased length of the smooth signal and HDO causes a higher BER, and on the contrary, a smaller BER close to original OFDM is achieved.
Finally, the analyzed error variance indicated that the training (reference) symbol is an effective alternative of time synchronization in the low-interference scheme.
From the results we conclude that, if the above two parameters are set reasonably, the low-interference scheme is capable of achieving an elegant balance between the sidelobe suppression, BER, complexity, and PAPR performances.
The analysis in this paper can be readily extended to OFDM-based wireless communication systems using millimeter wave (mmWave) spectrum \cite{Ref35}. 
In our future work we will investigate invoking the proposed measures of spectrum and error performance in order to enhance the spectral efficiency while maintaining the required transmission integrity.


\appendices
\section{Derivation of \eqref{Eqn:Eqn48}}

Substituting \eqref{Eqn:Eqn47} into \eqref{Eqn:Eqn46} yields
\begin{align}  \label{Eqn:Eqn72}
  p_{\rm b}({\rm E})=
  &  \frac{2}{\sqrt{J}\log_2\sqrt{J}} \sum\limits^{\log_2\sqrt{J}}_{u_1=1} \sum\limits^{(1-2^{-u_1})\sqrt{J}-1}_{u_2=0} (-1)^{\lfloor \frac{u_2 2^{u_1-1}}{\sqrt{J}} \rfloor}  \nonumber \\
& \cdot   \left( 2^{u_1-1}-\left \lfloor \frac{u_2 2^{u_1-1}}{\sqrt{J}} +\frac{1}{2}  \right \rfloor \right) I_1,
\end{align}
where
\begin{align}
  E_1=& \int\limits^{\sigma^{-}}_{0}   Q\left( (2u_2+1) \sqrt{\frac{3\gamma}{J-1}}  \right)    \frac{ \sigma^2_{\rm n}/M }{E\{\alpha\} (1-2\sigma^2_{\rm w}\gamma)^2}    \nonumber \\
& \cdot    e^{-\frac{\sigma^2_{\rm n}\gamma /M}{E\{\alpha\}(1-2\sigma^2_{\rm w}\gamma)}}  d\gamma.
  \label{Eqn:Eqn73}
\end{align}

Similar to \cite{Ref31}, by using $Q(x)=\frac{1}{2}(1-{\rm erf}(\frac{x}{\sqrt{2}}))$ and the error function's Maclaurin series ${\rm erf}(x)=\frac{2}{\sqrt{\pi}}\sum\limits^{+\infty}_{v=0}\frac{(-1)^v x^{2v+1}}{v! (2v+1)}$, we have
\begin{align}
  E_1&= \int\limits^{\sigma^{-}}_{0} \frac{1}{2}\left(1-{\rm erf}\left( (2u_2+1)\sqrt{\frac{1.5\gamma}{J-1}} \right)\right)   \nonumber \\
 &\quad \cdot  \frac{ \sigma^2_{\rm n}/M}{E\{\alpha\} (1-2\sigma^2_{\rm w}\gamma)^2} e^{-\frac{\sigma^2_{\rm n}\gamma/M}{E\{\alpha\}(1-2\sigma^2_{\rm w}\gamma)}}  d\gamma \nonumber \\
&=\frac{1}{2}-\frac{1}{2}\int\limits^{\sigma^{-}}_{0} {\rm erf}\left( (2u_2+1)\sqrt{\frac{1.5\gamma}{J-1}} \right) \nonumber \\
 &\quad \cdot \frac{ \sigma^2_{\rm n}/M }{E\{\alpha\} (1-2\sigma^2_{\rm w}\gamma)^2} e^{-\frac{\sigma^2_{\rm n}\gamma/M}{E\{\alpha\}(1-2\sigma^2_{\rm w}\gamma)}}  d\gamma \nonumber \\
&=\frac{1}{2}\!-\!\frac{1}{\sqrt{\pi}}\!  \sum\limits^{+\infty}_{v_1=0} \frac{(-1)^{v_1}  \sigma^2_{\rm n}/M}{v_1! (2v_1\!+\!1)E\{\alpha\}} \! \left(\frac{3(2u_2\!+\!1)^2}{2(J-1)}\right)^{\!\!v_1+\frac{1}{2}} \! E_2,
  \label{Eqn:Eqn74}
\end{align}
with
\begin{equation}
E_2=\int\limits^{\sigma^{-}}_{0} \frac{\gamma^{v_1+\frac{1}{2}}}{(1-2\sigma^2_{\rm w}\gamma)^2} e^{-\frac{\sigma^2_{\rm n}\gamma/M}{E\{\alpha\}(1-2\sigma^2_{\rm w}\gamma)}}  d\gamma.
  \label{Eqn:Eqn75}
\end{equation}

Then, using the Taylor series expansion of $e^x$ in \eqref{Eqn:Eqn75}, we arrive at
\begin{align}
E_2=\sum\limits^{+\infty}_{v_2=0} \frac{(-\sigma^2_{\rm n}/M)^{v_2}}{v_2!(E\{\alpha\})^{v_2}}
\int\limits^{\sigma^{-}}_{0} \frac{\gamma^{v_1+v_2+\frac{1}{2}}}{(1-2\sigma^2_{\rm w}\gamma)^{v_2+2}} d\gamma.
  \label{Eqn:Eqn76}
\end{align}

Furthermore, conditioned on $|\arg\{1-2\sigma^2_{\rm w}\sigma^{-}\}|<\pi$ and ${\rm Re}\{v_1+v_2+\frac{3}{2}\}>0$ \cite{Ref32}, we have
\begin{align}
& \int\limits^{\sigma^{-}}_{0} \frac{\gamma^{v_1+v_2+\frac{1}{2}}}{(1-2\sigma^2_{\rm w}\gamma)^{v_2+2}} d\gamma
=\frac{(\sigma^{-})^{v_1+v_2+\frac{3}{2}}}{v_1+v_2+\frac{3}{2}} \nonumber \\
& \quad \cdot {}_2F_1(v_2+2,v_1+v_2+\frac{3}{2};v_1+v_2+\frac{5}{2};2\sigma^2_{\rm w}\sigma^{-}).
  \label{Eqn:Eqn77}
\end{align}

Finally, upon substituting \eqref{Eqn:Eqn74}, \eqref{Eqn:Eqn76}, and \eqref{Eqn:Eqn77} into \eqref{Eqn:Eqn49}, we can obtain the BER expression in \eqref{Eqn:Eqn48}.

\section{Derivation of \eqref{Eqn:Eqn57}}

We first assume that the channel responses satisfy $h_{i,\tilde{l}}=h_{i+1,\tilde{l}}$ in the $(i+1)$th CP
overlapped by the $i$th channel-delayed symbol.
Then, upon substituting \eqref{Eqn:Eqn12} and \eqref{Eqn:Eqn51}-\eqref{Eqn:Eqn56} into $\int\limits^{T_{\rm s}-\delta_1}_{0} \!\!\!\!\!\! E\!\left\{\!\left| \bar{z}_{i}(t)\right|^2\!\right\}\!dt$ and $\int\limits^{T_{\rm s}}_{T_{\rm s}-\delta_1} \!\!\!\!\!\!  E\left\{\left| \bar{z}_{i}(t)\right|^2\right\}$, we can arrive at
\begin{align}  \label{Eqn:Eqn78}
& \int\limits^{T_{\rm s}-\delta_1}_{0} \!\!\!\!\!\! E\!\left\{\!\left| \bar{z}_{i}(t)\right|^2\!\right\}\!dt  \nonumber\\
&\quad=\!\!\!\!\! \int\limits^{T_{\rm s}-\delta_1}_{0} \!\!\!\!\!\! E\!\left\{\!\left| u_{i}(t\!+\!\delta_1)\right|^2\!\right\}dt \!+\!\!\!\!\!  \int\limits^{T_{\rm s}-\delta_1}_{0} \!\!\!\!\!\! E\!\left\{\!\left| n_{i}(t\!+\!\delta_1)\right|^2\!\right\}dt  \nonumber \\
&\quad=  \sum\limits^{\tilde{L}}_{\tilde{l}=1} \! E\{|h_{i,\tilde{l}}|^2\} \! \sum\limits^{N}_{n=0} \sum\limits^{N}_{\bar{n}=0} \sum\limits^{K-1}_{r=0} \! (a_{nr}a^{*}_{\bar{n}r}\!+\!b_{nr}b^{*}_{\bar{n}r})   \nonumber \\
&\quad \quad \cdot \int\limits^{T_{\rm s}-\delta_1}_{0} \tilde{f}_n(t-\tau_{\tilde{l}}+\delta_1) \tilde{f}^{*}_{\bar{n}}(t-\tau_{\tilde{l}}+\delta_1) dt \nonumber \\
&\quad\quad + (T_{\rm s}-\delta_1) \sigma^2_{\rm n},
\end{align}
and
\begin{align}  \label{Eqn:Eqn79}
&\int\limits^{T_{\rm s}}_{T_{\rm s}-\delta_1} \!\!\!\!\!\!  E\left\{\left| \bar{z}_{i}(t)\right|^2\right\}  \nonumber \\
&=\!\!\! \int\limits^{T_{\rm s}}_{T_{\rm s}-\delta_1} \!\!\!\!\!\! E\Big\{\!\Big| z_{i}(t + \delta_{1} )- \hat{z}_{i}(t + \delta_{1} )+z_{i+1}(t+\delta_{1} -T) \nonumber \\
&\quad +u_{i+1}(t + \delta_{1} - T) \Big|^2\!\Big\}dt +\!\!\!\!\!  \int\limits^{T_{\rm s}}_{T_{\rm s}-\delta_1} \!\!\!\!\!\! E\!\left\{\!\left| n_{i+1}(t+\delta_{1} -T)\right|^2\!\right\}dt  \nonumber \\
&= \delta_1  \sigma^2_{\rm n} + 2 K \sum\limits^{L_{1}}_{\tilde{l}=1}  E\{|h_{i,\tilde{l}}|^2\}  \left(\delta_1- \tau_{\tilde{l}} \right)  - \sum\limits^{L_{1}}_{\tilde{l}=1}  \sum\limits^{K-1}_{r=0} E\{|h_{i,\tilde{l}}|^2\}     \nonumber \\
&\quad \cdot \Bigg(  \!   2{\rm Re}\Bigg\{ \! \sum\limits^{N}_{n=0} b_{nr} \!\!\!\!\!  \int\limits^{T_{\rm s}}_{T_{\rm s}-\delta_1+\tau_{\tilde{l}}}  \!\!\!\!\!\!\!\!  e^{-j2\pi\frac{k_r}{T_{\rm s}}(t\!+\!\delta_1\!-\!T\!-\!\tau_{\tilde{l}})} \tilde{f}_n(t\!+\!\delta_1\!-\!T\!-\!\tau_{\tilde{l}}) dt  \Bigg\}   \nonumber \\
&\quad +   2 {\rm Re}\Bigg\{ \! \sum\limits^{N}_{n=0} a_{nr} e^{-j2\pi \frac{\delta_2}{T_{\rm s}}T}  \nonumber \\
&\quad  \cdot  \int\limits^{T_{\rm s}}_{T_{\rm s}-\delta_1+\tau_{\tilde{l}}}  \!\!\!\!\!\!\!\!  e^{-j2\pi\frac{k_r}{T_{\rm s}}(t\!+\!\delta_1\!-\!T_{\rm s}\!-\!\tau_{\tilde{l}})} \tilde{f}_n(t\!+\!\delta_1\!-\!T\!-\!\tau_{\tilde{l}}) dt  \Bigg\}    \nonumber \\
&\quad- \sum\limits^{N}_{n=0} \sum\limits^{N}_{\bar{n}=0} (a_{nr}a^{*}_{\bar{n}r}\!+\!b_{nr}b^{*}_{\bar{n}r})   \nonumber \\
& \quad \cdot \int\limits^{T_{\rm s}}_{T_{\rm s}-\delta_1+\tau_{\tilde{l}}} \!\!\!\!\! \tilde{f}_n(t\!+\!\delta_1\!-\!T\!-\!\tau_{\tilde{l}}) \tilde{f}^{*}_{\bar{n}}(t\!+\!\delta_1\!-\!T\!-\!\tau_{\tilde{l}}) dt   \Bigg).
\end{align}

Then, upon adding \eqref{Eqn:Eqn78} and \eqref{Eqn:Eqn79}, we can obtain $I_1 +\sigma^2_{\rm n} T_{\rm s}$, which is directly used to achieve \eqref{Eqn:Eqn57}.

\section{Derivation of \eqref{Eqn:Eqn66}}

We first assume that the channel responses satisfy $h_{i-1,\tilde{l}}=h_{i,\tilde{l}}$ in the $i$th CP
overlapped by the $(i-1)$th channel-delayed symbol.
Then, according to \eqref{Eqn:Eqn10}, \eqref{Eqn:Eqn12}, \eqref{Eqn:Eqn51}, \eqref{Eqn:Eqn52},  and \eqref{Eqn:Eqn63}-\eqref{Eqn:Eqn65}, the energy of the interference $\tilde{z}_i(t)$ in $[0,\delta_1\!+\! \tau_{\tilde{L}} \!-\! T_{\rm cp})$ and $[\delta_1\!+\! \tau_{\tilde{L}} \!-\! T_{\rm cp}, T_{\rm s}]$ is, respectively, derived as 
\begin{align}  \label{Eqn:Eqn80}
&\int\limits^{\delta_1\!+\! \tau_{\tilde{L}} \!-\! T_{\rm cp}}_{0} \! \! \!\! \!  \!\! \!  E\left\{\left| \tilde{z}_i(t) \right|^2\right\}dt \nonumber \\
&= \! \! \!\! \! \! \!   \int\limits^{\delta_1\!+\! \tau_{\tilde{L}} \!-\! T_{\rm cp}}_{0} \!\! \!  \! \! \! \!\! \!  E \! \left\{ \! \left| z_i(t\!-\!\delta_1) \!-\! \hat{z}_i(t\!-\!\delta_1) \!+\! u_i(t\!-\!\delta_{1})  \!+\! z_{i\!-\!1}(t\!+\!T\!-\!\delta_1) \right|^2 \! \right\} \! dt \nonumber \\
&=  2 K \!\!\! \sum\limits^{\tilde{L}}_{\tilde{l}=\tilde{L}-L_{2}+1} \!\!\!\!  E\{|h_{i,\tilde{l}}|^2\}  ( \tau_{\tilde{l}}\!-\!T_{\rm cp}+\delta_1)    \nonumber \\
&\quad + \!\! \sum\limits^{\tilde{L}}_{\tilde{l}=\tilde{L}-L_{2}+1}  \!\sum\limits^{K-1}_{r=0} E\{|h_{i,\tilde{l}}|^2\}      \nonumber \\
&\quad \cdot \sum\limits^{N}_{n=0} \sum\limits^{N}_{\bar{n}=0} (a_{nr}a^{*}_{\bar{n}r}\!+\!b_{nr}b^{*}_{\bar{n}r}) \!\!\!\!\! \!\!\!\!\!  \int\limits^{\delta_1\!+\! \tau_{\tilde{L}} \!-\! T_{\rm cp}}_{\delta_1\!+\! \tau_{\tilde{l}} \!-\! T_{\rm cp}} \!\!\!\!\! \!\!\!\!\!  \tilde{f}_n(t\!-\!\delta_1\!-\!\tau_{\tilde{l}}) \tilde{f}^{*}_{\bar{n}}(t\!-\!\delta_1\!-\!\tau_{\tilde{l}}) dt     \nonumber \\
&\quad +  \sum\limits^{\tilde{L}-L_{2}}_{\tilde{l}=1}  \!\sum\limits^{K-1}_{r=0} E\{|h_{i,\tilde{l}}|^2\}      \nonumber \\
&\quad \cdot \sum\limits^{N}_{n=0} \sum\limits^{N}_{\bar{n}=0} (a_{nr}a^{*}_{\bar{n}r}\!+\!b_{nr}b^{*}_{\bar{n}r}) \!\!\!\!\! \!\!\!\!\!  \int\limits^{\delta_1\!+\! \tau_{\tilde{L}} \!-\! T_{\rm cp}}_{0} \!\!\!\!\! \!\!\!\!\!  \tilde{f}_n(t\!-\!\delta_1\!-\!\tau_{\tilde{l}}) \tilde{f}^{*}_{\bar{n}}(t\!-\!\delta_1\!-\!\tau_{\tilde{l}}) dt, 
 \end{align}
and
\begin{align}  \label{Eqn:Eqn81}
&\int\limits^{T_{\rm s}}_{\delta_1\!+\! \tau_{\tilde{L}} \!-\! T_{\rm cp}}   \! \! \!\! \! \! \! \! E\left\{\left| \tilde{z}_i(t) \right|^2\right\}dt  \nonumber \\
&= \sum\limits^{\tilde{L}}_{\tilde{l}=1}  \sum\limits^{K-1}_{r=0} E\{|h_{i,\tilde{l}}|^2\}  \sum\limits^{N}_{n=0} \sum\limits^{N}_{\bar{n}=0} (a_{nr}a^{*}_{\bar{n}r}\!+\!b_{nr}b^{*}_{\bar{n}r}) \nonumber \\  
&\quad \cdot \int\limits^{T_{\rm s}}_{\delta_1\!+\! \tau_{\tilde{L}} \!-\! T_{\rm cp}} \!\!\!\!\! \!\!\!\!\!  \tilde{f}_n(t\!-\!\delta_1\!-\!\tau_{\tilde{l}}) \tilde{f}^{*}_{\bar{n}}(t\!-\!\delta_1\!-\!\tau_{\tilde{l}}) dt.
 \end{align}
 
Finally, adding \eqref{Eqn:Eqn80} and \eqref{Eqn:Eqn81} yields $I_3$ in \eqref{Eqn:Eqn67} to obtain \eqref{Eqn:Eqn66}.

\end{document}